\documentclass{aastex63}
\usepackage[T1]{fontenc}
	\usepackage{ae,aecompl}

\usepackage[toc,page]{appendix}

	\usepackage{soul}
	\usepackage[utf8]{inputenc}
	\usepackage{multirow}
	\usepackage{verbatim}
	\usepackage{color}

	\newcommand\s{\small}
	
	\newcommand\fns{\footnotesize}

	\usepackage{graphicx}

\begin{document}
			
			\title{On the Effects of Local Environment on
				AGN in the Horizon Run 5 Simulation}
            \shorttitle{Environmental Effects on AGN in HR5}
			\shortauthors{Singh et~al.}
		
			\correspondingauthor{Ena Choi}
			\email{enachoi@uos.ac.kr}
			
			\author[0000-0001-5427-4515]{Ankit Singh}
			\affiliation{Korea Institute for Advanced Study (KIAS), 85 Hoegiro, Dongdaemun-gu, Seoul 02455, Republic of Korea}
			
			\author[0000-0001-9521-6397]{Changbom Park}
			\affiliation{Korea Institute for Advanced Study (KIAS), 85 Hoegiro, Dongdaemun-gu, Seoul 02455, Republic of Korea}
			
			\author[0000-0002-8131-6378]{Ena Choi}
			\affiliation{Department of Physics, University of Seoul, 163 Seoulsiripdaero, Dongdaemun-gu, Seoul 02504, Republic of Korea}
			\affiliation{Korea Institute for Advanced Study (KIAS), 85 Hoegiro, Dongdaemun-gu, Seoul 02455, Republic of Korea}
			
			\author[0000-0002-4391-2275]{Juhan Kim}
			\affiliation{Korea Institute for Advanced Study (KIAS), 85 Hoegiro, Dongdaemun-gu, Seoul 02455, Republic of Korea}
			
			\author{Hyunsung Jun}
			\affiliation{Seoul National University (SNU), 1 Gwanak-ro, Gwanak-gu, Seoul 08826, Republic of Korea}

			\author[0000-0003-4446-3130]{Brad K. Gibson}
			\affiliation{E.A. Milne Centre for Astrophysics, University of Hull, Hull, HU6~7RX, United Kingdom}
			
			\author[0000-0003-4164-5414]{Yonghwi Kim}
            \affiliation{Department of Astronomy, Yonsei University, 50 Yonsei-ro, Seodaemun-gu, Seoul 03722, Republic of Korea}		
            \affiliation{Korea Institute for Advanced Study (KIAS), 85 Hoegiro, Dongdaemun-gu, Seoul 02455, Republic of Korea}

			\author[0000-0002-6810-1778]{Jaehyun Lee}
			\affiliation{Korea Institute for Advanced Study (KIAS), 85 Hoegiro, Dongdaemun-gu, Seoul 02455, Republic of Korea}

			\author{Owain Snaith}
			\affiliation{GEPI, Observatoire de Paris, PSL Universit\'e, CNRS, 5 Place Jules Janssen, 92190, Meudon, France}

\begin{abstract} 
			We use the Horizon Run 5 cosmological
			simulation to study the effect of galaxy intrinsic properties and the local environment on AGNs characterized by their threshold of the accretion rate. We select galaxies in the stellar mass range $10^{9.5} \le M^{}_{*}/M^{}_{\odot} \le 10^{10.5}$ in the snapshot at redshift $z$=0.625. Among various intrinsic properties, we find that the star formation rate of the host galaxy is most correlated to the AGN activity. To quantify the environment, we use background galaxy number density (large-scale environment) and 
			distance and morphological type of the nearest neighbors (small-scale environment),
			and study their relative effects on the AGN properties. We find that, compared to the background density, 
			the nearest neighbor environment is the dominant 
			quantity determining the bolometric luminosity, star formation rate, and kinematic properties of AGNs and better dictates the gas mass of the host galaxy. We show that the cold gas content in the host galaxies is crucial in triggering AGN activity.
			However, when the nearest neighbor environment effects start to act at the neighbor distance of less than about half the virial radius of the neighbor, the neighbor environmental effects are the most dominant factor for quasar activity.
			\end{abstract}
			

\section{Introduction}\label{sec:intro}

Observations have shown that the star formation rate density in the Universe has declined in the current epoch  \citep{2004Hopkins,2013Behroozi,2014Madau}. One of the key mechanisms proposed for the quenching of massive galaxies is feedback from the Active Galactic Nucleus \citep[AGN][]{2005Matteo,2006Bower,2007Schawinski,2012Cano,2012Maiolino,2012Fabian,2012Dubois,2013Dubois,2017Beckmann,2021Zhang}. The gas inside a galaxy loses the angular momentum and falls towards the supermassive black hole residing inside the galaxy. The energy released from the accretion heats the surrounding gas and is thought to quench the star formation \citep{2010Antonuccio,2013Wagner}. One of the ways to classify AGNs as quasar mode \citep[radiationally efficient;][]{1973Shakura} and radio mode \citep[radiationally inefficient;][]{1979Hine} is based on the rate of accretion. \footnote{Radio mode does not necessarily involve observed radio emission, but we are simply comparing high and low-accretion AGNs.} Quasar mode AGNs are associated with actively star-forming disk galaxies, and radio mode is mostly associated with passively star-forming elliptical galaxies \citep[][and references within]{2014Heckman}. The trigger mechanisms of AGN have been a subject of debate since their discovery.\par

It has been suggested that AGNs can be triggered by internal and external factors. Internal properties like gas content of the host galaxy, kinematics, and morphology could potentially play a role in regulating the accretion of gas onto the supermassive black hole at the center \citep[for example;][]{2010Gavignaud,2014Dubois,2019Ruffa,2020Shangguan,2021Ellison}. The environment of a galaxy, e.g., how many galaxies it has in its vicinity, on small and large scales, has been observed to play a critical role in regulating the galaxy properties like star formation rate, gas content, and stellar and gas metallicity \citep{1988Binggeli,1997Dressler,2004Kauffmann,2008Porter,2009Weinmann,2010Mahajan,2012Peng,2015Alpaslan,2017Kuutma,2018Mahajan,2020Asano,2020Singh,2021Gouin}. Galaxies can acquire gas from its environment fueling AGNs \citep{2006Allen,2007Hardcastle,2015Ineson}. Similarly, interaction with the neighbors has been shown to regulate the gas content and morphology of the galaxies undergoing interaction. The gravitational interaction can funnel the gas toward the center of a galaxy \citep{1989Hernquist,1992Barnes,2008Hopkins,2015Comerford,2021Sharma}. Therefore, it is important to study the effect of these external processes on the AGNs.\par

The prevalence of AGNs in different environments has been studied extensively over the past decade \citep{2007Gilmour,2011Bradshaw,2012Hwang,2015Malavasi}.
\cite{2010Padilla} showed that at fixed background density, AGN host
galaxies are relatively redder compared to similar galaxies with weak or no AGN, which hints at an environmental effect on AGNs. \cite{2014Donoso} reported that obscured AGNs are preferably found in the denser environment at $z\sim1$. \cite{2016Argudo} used optical and radio AGNs to report
that optical AGNs are not affected by background density, whereas radio AGNs are
strongly affected by it. At higher redshifts ($1.4 \leq z \leq 2.5$), \cite{2020Bornancini} studied obscured and unobscured AGNs and QSOs, reporting the correlation of different AGN types with the environment. \cite{2021Santos} using AKARI North Ecliptic Pole Wide field reported that at redshift ($0.7 \leq z \leq 1.2$), AGN activity increases with an increase in environmental density. \par

\cite{2011Ellison} used SDSS to show that the fraction of AGN galaxy with a close companion is higher compared to a control sample and concluded that interaction plays a role in triggering AGN activity. \cite{2015Hong} found that 17 of 39 AGNs at low redshift showed signs of mergers in the past. \cite{2012Treister} used multi-wavelength observations of AGNs in redshift range $(0 < z < 3)$ to study relation between AGNs and major mergers. They concluded that the most luminous AGNs are driven by major mergers, and less luminous AGNs are powered by secular processes. \cite{2014Satyapal} found that the fraction of AGNs increases as the distance to neighbors decreases, and mergers can enhance the AGN activity. Recently, \cite{2021Zhang} using SDSS showed that AGNs are mostly found in starburst and green-valley phases. They found that AGNs are surrounded by more neighbors compared to star forming galaxies. The results support the scenario that interaction could trigger AGN activity. \par

Hydrodynamical cosmological simulations have been used to model the effects of AGNs and compare them with observations. \cite{2005Matteo} performed merger simulations to study the role of mergers in triggering AGNs. They concluded that mergers apart from intense star formation also lead to gas inflow to the central black holes, which can power the quasars. \cite{2020Bhowmick} used IllustrisTNG simulation \citep{2015Nelson,2018Pillepich} to study the AGN environment within 0.01-1 $h^{-1}$ Mpc and found that interaction increases the AGN activity but plays a minor role. \cite{2021Kristensen} used TNG100-1 run (with box size 75 $\rm Mpc \  h^{-1}$, $1820^3$ dark matter particles with particle mass resolution of $ \approx 10^{6} \ M_{\odot}$) to study the environment of dwarf AGN host galaxies. They found intermediate AGN activity in the galaxies that experienced recent minor mergers. \par

It is clear from the discussion above that simulations, like the observations, have different conclusions on what role galaxy properties and environment play in dictating the activity of the central black holes. For example, is the aggregated effect of many galaxy interactions, like in a group environment, more important? Or does one-to-one interaction matter for AGNs? Furthermore, How the activity of the central black hole relate to the intrinsic host galaxy properties like star formation rate, gas mass, and metallicity? In this study, we try to answer these critical questions by performing a statistical study of AGN activity with galaxy properties and the local environment.

Horizon Run 5 simulation \citep[HR5;][]{2020Lee,Park2022} has a large box with advanced numerical techniques that allow us to have a high enough number of galaxies in different modes of AGN activity. HR5 enables us to probe the AGNs with a good variation in evolutionary history and environments. In Section \ref{sec:Method} we outline the methodology used in the paper. We present our results in Section \ref{sec:results} followed by a discussion in Section \ref{sec:diss}.

\section{Method}\label{sec:Method}

In the following subsections, we will briefly describe the details of HR5
simulation (\ref{sec:hr5}). We refer the reader to \cite{2020Lee} for further
details. We will describe our selected sample for the study in
\ref{sec:sample}. We describe morphology, local environment and neighbor selection in Sections \ref{sec:morpho}, \ref{sec:env} and \ref{sec:nn} respectively.

\subsection{Horizon run 5 simulation}\label{sec:hr5}

Horizon Run 5 (HR5) is a  hydrodynamical simulation run using
Adaptive Mesh Refinement (AMR) code \citep[\texttt{RAMSES}; ][]{2002Teyssier}. The box size
of HR5 is \( \rm (1049 \  cMpc )^3\) with zoom-in region of  \( \rm 1049 \times 127 \times 119 \ cMpc^3\)  crossing the center of the box. The grids are refined to an AMR level of 20 to achieve a spatial resolution down to 1 kpc in a zoomed region. The initial condition of the
simulation uses the parameters compatible with those of Planck \citep{Planck} for cosmology ($\Omega_m = 0.3$, $\Omega_{\Lambda} = 0.7$, $\Omega_b =0.047$, $\sigma_8 = 0.816$, and $h_0=0.684$) with a linear
power spectrum generated using the CAMB package \citep{2000Lewis}. \par

\subsubsection{Physical processes}\label{sec:phy_params}

The gas is allowed to cool down to a temperature of $10^4$ K  using cooling functions
proposed by \cite{1993Sutherland}. The metal-rich gas is further allowed to
cool down to a temperature of 750 K using cooling rates given by
\cite{1972Dalgarno} in the presence of uniform UV background \citep{1996Haardt}
to mimic reionization at $z=10$. \par 

The star formation up to $z=21$ follows an approach similar to \cite{2006Rasera}
in which stars are formed in the cells with gas number density
$n_{g}>55\rho_{\rm crit}$, where $\rho_{\rm crit}$ is the critical density of
the universe. After $z=21$, the star-formation criterion is set to
$n_{g}>0.1 \  \rm H \ cm^{-3}$ (Hydrogen atoms per cubic centimeter) for the cells with the temperature lower than
2000 K. The assigned mass of the stellar particles is given by:

\begin{equation} m_* =
	0.2N_* (f_{\rm b}\Delta x_{\rm res}^3) \,,
\end{equation}
 where $f_{\rm b}=\frac{\Omega_\mathrm{b}}{\Omega_{m}}$ and $\Delta x_{\rm res}$ is the spacial resolution at highest refinement level. $N_{*}$ is a random variable picked up from a Poisson distribution, and the $m_*$ is mass in code units (total mass of matter set equal to unity). Based on a Schmidt law \citep{1959Schmidt}  the star formation rate is given by:

\begin{equation} \dot{\rho_*} = \epsilon_* \rho_\mathrm{g} \sqrt{\frac{32 G
			\rho_\mathrm{g}}{3 \pi}}\,, 
\end{equation} 
 where  $\rho_{\rm g}$ is the local gas density in the cell marked for star formation, $\dot{\rho_*}$ is the
star formation rate, and $\epsilon_* (=0.02)$ is the star
formation efficiency, and $G$ is the gravitational constant. In order to prevent excessive gas depletion, no more than
90 \% of the gas in the cell is allowed to convert into stars.

\subsubsection{Feedback processes} \label{sec:feedback}

HR5 implements various stellar feedback recipes that mimic
winds from the Asymptotic Giant Branch (AGB) stars, supernova type-Ia, and
type-II. The energy released per supernova event is set to $2\times10^{51}$
ergs. Young particles contribute to the supernova type-II feedback by depositing
a certain fraction of their energies in kinetic ($f_{\rm k}=0.3$) and rest as
thermal energy. Ejected mass and energy are deposited in a radius of 2 times
the one-dimensional size of the most refined gas cells ($\rm \sim 1 \ kpc$). In the aged stellar
particles, the feedback is continued in the form of supernova type-Ia and AGB
winds in the form of thermal deposition in nearby cells.\par

In the region with gas density $n_{\rm H,0} \geq 0.1 \ \rm H \ cm^{-3}$ and without black holes (BHs) within 50 kpc, a BH is created with a seed mass of $\mathrm{10^4 \ M_{\odot}}$ at the center of the gas cell of interest. BHs experience a drag force \citep{2014Dubois}. BHs are allowed to coalesce when the separation between them becomes less than $4\Delta$x, where $\Delta x$ is the size of the cell. BHs grow
by gas accretion following boosted  Bondi-Hoyle-Lyttleton
accretion rate \citep{1939Hoyle,1944Bondi,1952Bondi,2009Booth} given by:

\begin{equation} 
	\dot M_{\rm BH}=(1-\epsilon_{\rm r})\dot M_{\rm BHL} \ ; \  \dot M_{\rm BHL}=\frac{\alpha 4\pi\bar \rho G^2M_{\rm BH}^2}{(\bar u^2+ \bar c_{\rm s})^{3/2}}\,,
\end{equation} 
 where
$\epsilon_{\rm r}$ is the spin-dependent radiative efficiency
\citep{2014Dubois}, $\bar c_{\rm s}$ and $\bar u$ are the kernel-weighted sound speed and local gas velocity respectively, $\bar \rho$ is the kernel-weighted gas density of the local
medium, and $\alpha$ is a dimensionless boost factor ($\ge 1$). The accretion rate is capped at the Eddington accretion rate given by $\dot M_{\rm Edd}=4\pi G M_{\rm BH}m_{\rm p}/(\epsilon_{\rm r}\sigma_{\rm T} c)$ where $\sigma_{\rm T}$
is the Thomson cross-section and $v$ is the velocity of the BH relative to the
ambient medium, $m_{\rm p}$ is the mass of a proton, $c$ is the speed of light in vacuum. Depending on the Eddington ratio ($\chi=\dot M_{\rm BH}/\dot M_{\rm Edd}$) the feedback is delivered in two modes \citep{2008Merloni,2012Dubois}: quasar ($\chi > 0.01$) and radio ($\chi \le 0.01$). \par

In quasar mode, a geometrically thin, radiatively efficient disk is assumed
\citep{1973Shakura}. The gas is ejected isotropically in a sphere with an energy ejection rate given by:

\begin{equation} 
	\dot E_{\rm BH,h}=\epsilon_{\rm r}\epsilon_{\rm f,h}\dot M_{\rm BHL} c^2 \,,
\end{equation} 
where $\epsilon_{\rm f,h}$ is the
thermal efficiency set to 0.15. In the radio mode, BH spin evolution is assumed
to have magnetically choked accretion flow solution \citep{2012McKinney}.
The energy ejection rate is given by:

\begin{equation} 
	\dot E_{\rm BH,j}=\epsilon_{\rm f,j}\dot M_{\rm BHL} c^2\,,
\end{equation} 
with the thermal coupling efficiency ($\epsilon_{\rm f,j}$) distribution has
a U-shape with a maximum of 100\% for maximally spinning BHs and $~0.01$ for
non-rotating BHs \citep{2021Dubois}. This energy is injected into the
neighboring cells in a bipolar manner along the axis of the spinning BH with no
opening angle and a mass loading factor of $\eta=\dot M_{\rm J}/\dot M_{\rm BH}=100$, where $\dot M_{\rm J}$ is the rate of mass deposition.

\subsubsection{Galaxy identification}\label{sec:galinden}

To identify the FOF halos as usually dealt with in the standard halo finding methods in the snapshots of HR5, we treat the gas cells as `gas-particles' using mean density and level of refinement. These, along with stellar, dark matter, and BH, were used by extended
Friends-of-Friends \citep[FoF;][]{1982Huchra,1982Press,1985Davis} algorithm with variable linking length to identify virialized structures composed of multiple matter species. The linking length is given by:

\begin{equation} 
	l_{\rm link} = 0.2 \times \left( \frac{m_{par}}{\Omega_{m0} \rho_{\rm c}} \right)^{1/3} \,,
\end{equation}
 where $m_{\rm par}$ is the mass of a particle, $\rho_{\rm c}$ is the critical density of the Universe, $\Omega_{\rm m0}$ is the baryonic density parameter at $z=0$. For linking particles with different masses, average linking lengths are used. The virial radius of the neighbor galaxy is defined as $R_{\rm vir} = ( 3M_{\rm vir}/4\pi \rho_{200} )^{1/3}$, where $\rho_{200}$ is 200 times the critical density of the universe. \par 

Galaxies in the FoF halos have been identified using the PSB-based Galaxy Finder ($\texttt{pGalF}$) algorithm based on \cite{2006Kim}. Instead of dark matter, $\texttt{pGalF}$ uses stellar particles as they are more concentrated, making them better suited for identifying the internal structures of galaxies. The seeds from the stellar distribution are used for identifying structures of other components of the galaxy. We give a very brief introduction to $\texttt{pGalF}$ in the following paragraph. For details of the algorithm, we refer the reader to \cite{2020Lee}. $\texttt{pGalF}$ identifies galaxies in four steps:

\begin{itemize} 
	\item Create a network of nearest neighbors.  
	\item Identifying
	peaks of stellar density and identifying core particles. Core members are
	identified as the particles residing in a region identified by monotonically
	lowering the density threshold till the isodensity surface encloses another
	peak.  
	\item Grouping the non-core particles using the watershed algorithm.
	\item Membership check using tidal radius and total energy of particles in identified regions.  
	
\end{itemize}

\noindent A total of 158754 galaxies with stellar mass $M^{}_{*} \geq 2 \times 10^9 \ \rm M_{\odot}$ are
identified at the snapshot corresponding to $z=0.625$, the last snapshot of HR5. The galaxies are those at least 3 cMpc away from the regions containing low-level particles to avoid potential boundary effects.

\begin{figure} 
	
	\centering
	
	\includegraphics[width=0.45\textwidth]{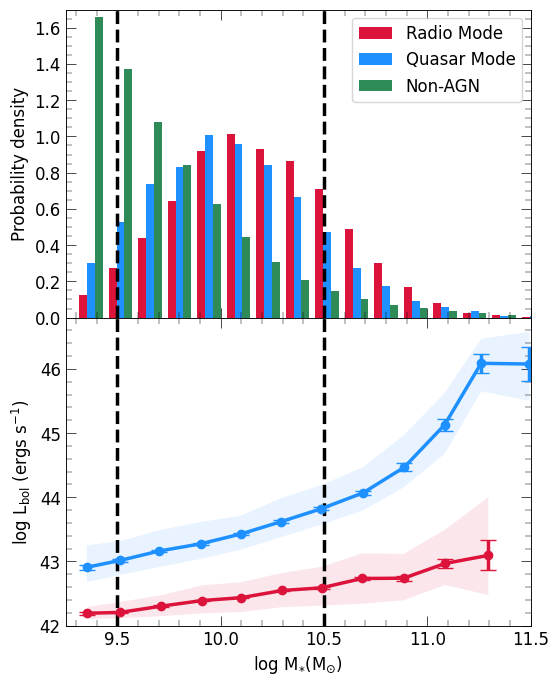}
	\caption{(top) Probability distribution function of stellar mass of the galaxies with $M^{}_{*} \geq \rm 2 \times 10^{9} \ M^{}_{\odot}$. The colors represent AGNs in radio (red) and quasar (blue) modes, with green indicating non-AGN galaxies. The distribution is similar in both modes suggesting galaxies hosting a radio mode AGN or a quasar mode AGN have a wide range of stellar mass. The black vertical dotted lines show the mass range of our selected sample $S_{\rm bin}$. (bottom) The bolometric AGN luminosity ($L_{\rm bol}$) is plotted against stellar mass ($M^{}_{*}$) 
	for the sample of galaxies $S_{\rm all}$.The bolometric luminosity of AGNs in both modes varies significantly with stellar mass. In the log bolometric luminosity varies linearly with log stellar mass in the range of $S_{\rm bin}$ is linear. We use the median bolometric luminosity for each stellar mass in this linear relation ($L_{\rm scale}$) to normalize the bolometric luminosity for each mode.} 
	
	\label{fig:mstar_dist} 
	
\end{figure}

\begin{figure} \centering
	
	\includegraphics[width=0.45\textwidth]{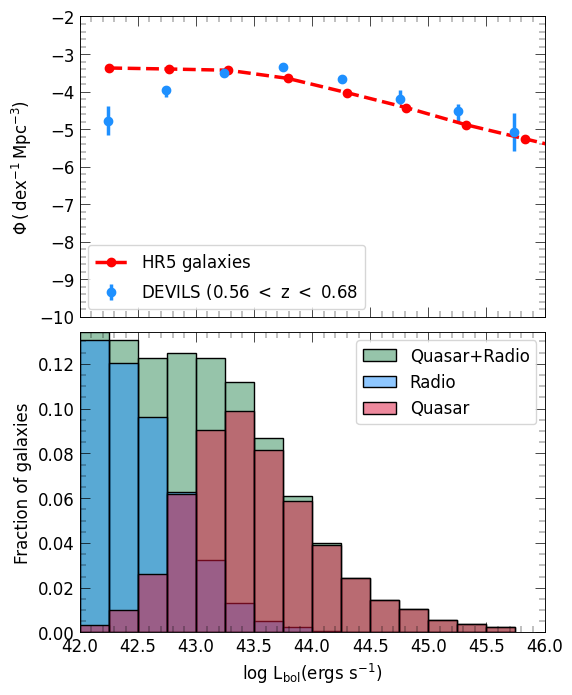}
	
	\caption{(top) The bolometric AGN luminosity function for the sample of galaxies $S_{\rm all}$. The blue points are from ~\protect\cite{2022Thorne} for the Deep Extragalactic VIsible Legacy Survey  ~\protect\citep[DEVILS;][]{2018Davies}. The HR5 bolometric luminosity agrees well with the observation within 0.8-1 dex except at the low luminosity which shows a discrepancy of $\sim 1.5$ dex. The difference at low luminosity can be attributed to the observational limitations in the detection of low-luminosity AGNs. (bottom) Normalized distribution of bolometric luminosity for HR5 simulation galaxies of the subsample $S_{\rm bin}$ for two modes of AGN. The quasar expectedly shows higher bolometric luminosity distribution than radio-mode galaxies. } \label{fig:Lbol_dist} 
	
\end{figure}

\begin{figure}
	
	\centering
	
	\includegraphics[width=0.45\textwidth]{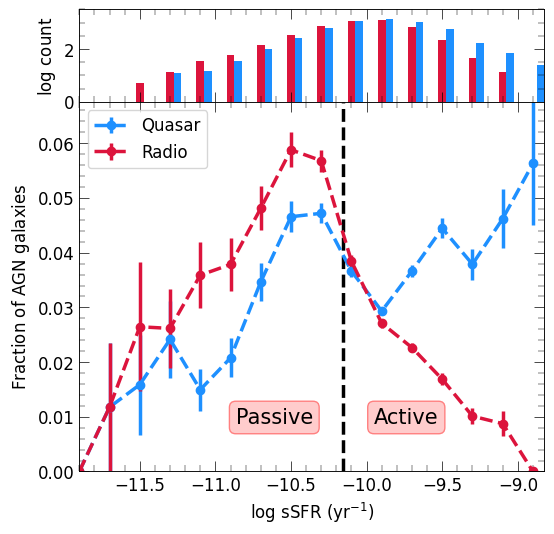}
	
	\caption{Fraction of radio and quasar mode galaxies in active and passive galaxies. The black vertical dotted line shows the value, $\rm sSFR_{\rm c} = 1/t_{\rm H0}$. Galaxies with specific star formation higher and lower than $\rm sSFR_{c}$ are classified as active and passive, respectively. Quasar mode is more prevalent in star-forming galaxies, whereas radio mode is dominant in passive galaxies. The histogram on top shows the number of AGN galaxies in each bin for the two modes.}
	
	\label{fig:AGNsSFR}
	
\end{figure}

\subsection{Galaxy sample \& variables}\label{sec:sample}

For our initial analysis, we make use of all the galaxies at redshift ($z$) of
0.625 in HR5 zoom-in region with stellar mass $M^{}_{*}$ greater than $2 \times 10^9 \ \rm M_{\odot}$. To calculate the bolometric luminosity of galaxies, we follow \cite{2019Griffin}, and \cite{2019Amarantidis}. \par

As explained earlier, the bolometric luminosity due to the accretion of gas is divided into two modes:  the quasar mode and the radio mode. In quasar mode,  the bolometric luminosity is given by $L^{\rm q}_{\rm bol}= \epsilon_{\rm r} \dot{M}_{\rm BH} c^2$, where $\epsilon_{\rm r}$ is the radiative efficiency, $\dot{M}_{\rm BH}$ is the accretion rate on the SMBH and $c$ is the speed of light \citep{1973Shakura}. In the case of the radio mode, the $L_{\rm bol}$ is dependent on an accretion threshold ($\chi_{\rm crit,\nu}$), denoting the regime in which the electrons are heated by the transfer of viscously generated energy from the ions \citep{1997Mahadevan}. The bolometric luminosity is given by:

\begin{equation}
	L_{\rm bol}=\left
	\{\begin{array}{l}
		\text{[if $\chi < \chi_{\rm crit,\nu}$]:}\\
		0.0002 L^{\rm q}_{\rm bol}\bigg(\frac{\delta}{0.0005}\bigg)\bigg(\frac{1-\beta}{0.5}\bigg)\bigg(\frac{6}{\hat{r}_{\rm lso}}\bigg),\\\\
		\text{[if $\chi_{\rm crit,\nu} \leq \chi < 0.01$]:}\\
		0.2 L^{\rm q}_{\rm bol}\bigg(\frac{\chi}{\alpha^2}\bigg)\bigg(\frac{\beta}{0.5}\bigg)\bigg(\frac{6}{\hat{r}_{\rm lso}}\bigg),
		\quad \\\\
		\text{[if $0.01 \leq \chi < \eta_{\rm edd}$]:}\\
		L^{\rm q}_{\rm bol},\\\\
		\text{[if $\chi \geq \eta_{\rm edd}$]:}\\
		\eta_{\rm edd}(1+\rm ln(\chi/\eta_{\rm edd}))L_{\rm edd}
	\end{array} \right.\,,
	\label{eq:Lbol}
\end{equation}
 where
\begin{equation}
	\chi_{\rm crit,\nu} = 0.001 \bigg(\frac{\delta}{0.0005}\bigg)\bigg(\frac{1-\beta}{\beta}\bigg)\alpha^2_{\rm radio}\,.
\end{equation}
Here $\hat{r}_{\rm lso}$ is the  last stable orbit around a SMBH in the units of gravitational radius ($GM_{\rm BH}/c^2$), $\alpha$ is the Shakura–Sunyaev viscosity parameter (set here to  0.1),  $\beta$ is the
fraction of gas pressure to total pressure ($= 1 - \alpha/0.55$),  $\delta$ is the fraction of energy received by electrons in viscous dissipation in accretion flow (set here to 0.0005), and $\eta_{\rm edd}$ is a free parameter which has been set to 4  \citep{2019Amarantidis}. This value gives similar luminosity in super-Eddington for a given mass accretion. This total sample of AGN galaxies is referred to as $S_{\rm all}$ in the following sections. The star formation rate ($SFR$) for galaxies equal to zero was given the value of $10^{-3} \rm M^{}_{\odot} yr^{-1}$. In Table \ref{tab:sample} we show the number of galaxies sufficiently away from the regions contaminated by low-level particles in each mode for both samples.\par

Figure \ref{fig:mstar_dist}(top) shows the probability distribution function of the stellar mass of the galaxies in our selected sample $S_{\rm all}$ for both modes and non-AGNs. We note that most galaxies in both modes are distributed in the stellar mass range of $10^{9.5} \le M^{}_{*}/M^{}_{\odot} \le 10^{10.5}$. In the bottom plot, the bolometric AGN luminosity ($L_{\rm bol}$) is plotted against stellar mass ($M^{}_{*}$). The bolometric luminosity of AGNs in both modes varies significantly with changes in stellar mass. The log bolometric luminosity varies linearly with log stellar mass in the range of  $10^{9.5} \le M^{}_{*}/M^{}_{\odot} \le 10^{10.5}$ is linear. We select this mass range as beyond this stellar mass range, the variation of bolometric luminosity with stellar mass becomes highly non-linear for the quasar mode and any secular evolution will be difficult to disentangle from the environmental effects. Note, that galaxies with a stellar mass more than $10^{10.5} \ M_{\odot}$ can be present at the center of the groups and clusters, which are known to host radio-mode galaxies, but for our study, these have been neglected.
Apart from the bolometric luminosities, other intrinsic properties also exhibit association with the stellar mass of galaxies. To study the relationship between AGN activity and properties of AGN host galaxies we have to remove the degeneracy with the secular variation due to the stellar mass. \par

We use the median bolometric luminosity for each stellar mass in this linear relation ($L_{\rm scale}$) to normalize the bolometric luminosity for each mode. We take AGNs with stellar mass range  $10^{9.5} \le M^{}_{*}/M^{}_{\odot} \le 10^{10.5}$ from the total $S_{\rm all}$ and refer to it as $S_{\rm bin}$ in the following sections. The selection retains a large number of galaxies for a statistical study. We have 8042 AGNs ($\sim 7$\%) of total galaxies in $S_{\rm bin}$. We scale the internal properties of AGN host galaxies (e.g. X) with the median value of those properties from a stellar mass-matched sample of Non-AGN galaxies (e.g. $X_{NonAGN}$) with stellar mass bins of 0.01 dex. The selection and scaling criterion help in neglecting the effects of the secular evolution of galaxies \citep{2004Kormendy} and study the effects of the environment and host galaxy properties on AGN activity.\par

In the top panel of Figure \ref{fig:Lbol_dist}, we show the bolometric AGN luminosity function ($\phi$) for galaxies in the sample $S_{\rm all}$ and compare it with the data from \cite{2022Thorne} for the Deep Extragalactic VIsible Legacy Survey  \citep[DEVILS;][]{2018Davies}. The red line shows the values for $L_{\rm bol}$ obtained from Equation \ref{eq:Lbol} for the HR5 galaxies and the blue points show the data from observations. Two agree well within 0.8-1 dex except at the low luminosity bin which shows a discrepancy of $\sim 1.5$ dex. The larger discrepancy at lower luminosity can be attributed to the observational limitations in the detection of low-luminosity AGNs. Given the limited size of the simulation box compared to the observation volume, it is good that the luminosity function matches well with observation for a significant luminosity range. Motivated by this we choose a cut-off of $L_{ \rm bol} \rm  \geq \rm 10^{42} \  ergs \ s^{-1}$ for studying the AGNs in HR5. In the bottom panel of Figure \ref{fig:Lbol_dist}, we show the bolometric luminosity distribution of the galaxies in the sample $S_{\rm bin}$ for both modes. As expected, the quasar-mode galaxies show a higher bolometric luminosity than radio-mode galaxies. Table \ref{tab:sample} shows the number of galaxies in each subsample we have taken for the study.

\begin{table} 
	\centering 
	\begin{tabular}{|c|c|c|c|} 
		\hline Sample & Total galaxies & Radio & Quasar \\ 
		\hline $S_{\rm all}$ & 158754 & 4792 & 5526  \\
		\hline $S_{\rm bin}$ & 100687 & 3621 &4421 \\ 
		\hline 
	\end{tabular}
	\caption{Total number of galaxies in each subsample we have selected. See the
		text for detail.  $S_{\rm all}$ is the total sample containing all the galaxies with stellar mass $M^{}_{*} \geq 2\times 10^{9} \ M_{\odot}$ and $S_{\rm bin}$ is the subsample of galaxies in the stellar mass range  $10^{9.5} \le M^{}_{*}/M^{}_{\odot} \le 10^{10.5}$ (see the text for description) used for analysis in the study. } 
	\label{tab:sample} 
\end{table}

\begin{figure} \centering
	
	\includegraphics[width=0.45\textwidth]{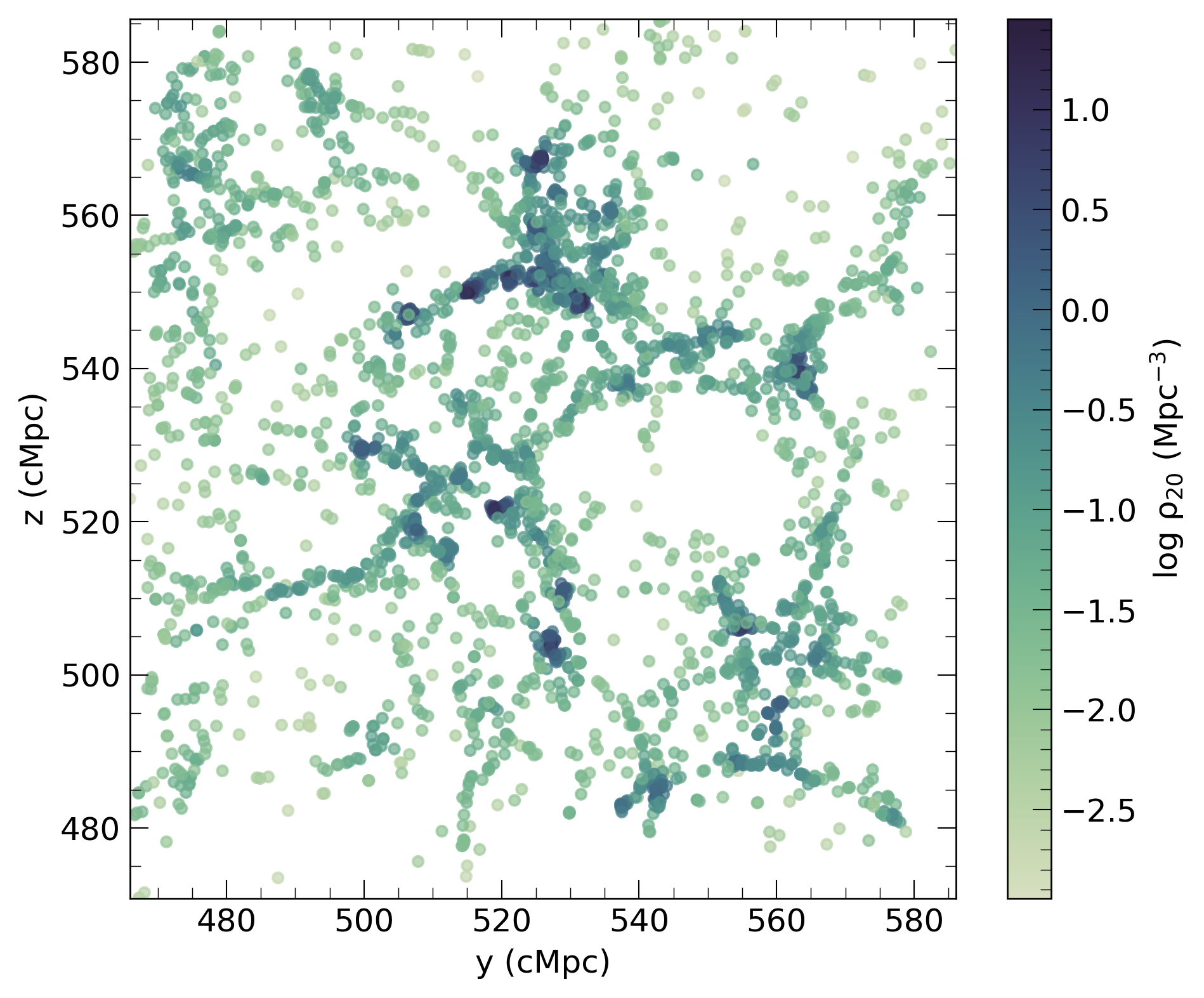}
	\caption{Background density at the locations of all HR5 galaxies with $M^{}_{*}/M^{}_{\odot} \geq 2 \times 10^{9}$ ($S_{\rm all}$) in a slice with thickness 25 cMpc. The colors show background density ($\rho_{20}$) obtained using an adaptive spline kernel smoothing (see text for details). The algorithm can capture the varying densities and complicated geometry in a good manner. The density decreases from dense groups and cluster environments to low-density void environments. }
	
	\label{fig:slice} 
	
\end{figure}

Galaxies can be classified as actively or passively star-forming based on their specific star formation rate (sSFR). In Figure \ref{fig:AGNsSFR} we divide the galaxies into active and passive depending on sSFR. The galaxies with $\rm sSFR<1/t_{\rm H0}$ and  $\rm sSFR \geq 1/t_{\rm H0}$ are classified as active and passive galaxies respectively, where $\rm t_{\rm H0}$ is the Hubble time. A value of $\rm sSFR<1/t_{\rm H0}$ means the time needed to form all stars in the galaxy was shorter than the Hubble time. It is clear that the radio mode AGNs are more prevalent in passive galaxies, and the quasar mode is dominant in the actively star-forming galaxies. \par

\subsection{Morphology of galaxies}\label{sec:morpho}
The morphology of galaxies is quantified using two
quantities: the S\'{e}rsic index \citep[$n$;][]{1963Sersic} and the asymmetry parameter ($A/A_{\rm NonAGN}$). The galaxies' stellar particles are projected face-on. The S\'{e}rsic index ($n$) of galaxies is calculated by fitting the radial distribution with the S\'{e}rsic profile given by:

\begin{equation}
	I(R) = I_{\rm e} \exp \biggl\{  \biggl(\frac{R}{R_{\rm e}}\biggr)^{1/n}\biggr\}\,
\end{equation}
\noindent
where $R$ is the projected radial distance from the center of a galaxy, $R_{\rm e}$
is the radius containing half the stellar mass of the galaxy and $I_{\rm e}$
is the intensity at this radius. $n$ is
called the  S\'{e}rsic index describing the shape of the profile.
The asymmetry parameter \citep[$A/A_{\rm NonAGN}$;][]{1995Schade,1996Abraham,2000Conselice,2003Conselice,2008Hernandez,2016Pawlik} is calculated in 3D using
the equation:

\begin{equation}
	A = \frac{\Sigma | (I_{\rm o} - I_\phi)|}{2\Sigma | I_{\rm o} |}\,,
\end{equation}
where $I_{\rm o}$ is the mass density of the original distribution of stellar
particles in 3D and $I_\phi$ is the density when a galaxy is inverted
by $180$ about the galaxy center. Most galaxies in the HR5 have a S\'{e}rsic index ($n$) of 1.25 with a range of 0.15-2.25 and the asymmetry parameter ($A/A_{\rm NonAGN}$) is in the range of 0.15-0.2.

\subsection{Local Environment}\label{sec:env}

To quantify the effect local environment, we use two parameters: distance to
the neighbor and background density estimate. \par 

For estimating the background density around the AGNs we use an adaptive spline-kernel smoothing with adapter-kernel size \citep{2015Huillier,2016Song}. The number density at a given position ($r_{\rm 0}$) is given by:
\begin{equation}
	\rho(r_{\rm 0}) = \sum_{i=1}^{N_\textrm{\s nn}} W(|r_{0} - r_i|,h_\textrm{\s spl})\,,
\end{equation}
where $W(|r_{0} - r_i|,h_\textrm{\s spl})$ is a Spline function kernel,
which depends on distance to the $i$-th nearest neighboring ($|r_{\rm 0} - r_{\rm i}|$), and a smoothing length ($h_{\rm spl}$) taken as half the distance to $N_{\rm nn}$th neighbor. Here we use 20 nearest neighbors, which is large enough to minimize the effect of short noise below 1\%  as demonstrated in \cite{2016Song}. The kernel used for density estimation should satisfy the properties of normalization and compactness. In this work, we use the Spline function kernel proposed by \cite{1985Monaghan}, given as:

\begin{equation}
	W(r_i,h_\textrm{\s spl})
	=\frac{1}{\pi h_\textrm{\fns spl}^3} 
	\begin{cases}
		1-\frac{3}{2}q_i^2 + \frac{3}{4}q_i^3 \,; \quad q_i\le1\\
		\frac{1}{4}(2-q_i)^3 \,; \quad 1\le q_i\le2\\
		0 \,; \qquad\qquad \text{otherwise}
	\end{cases}\,,
\end{equation}
where $q_i=r_i/h_\textrm{\s spl}$. 

In Figure \ref{fig:slice} we show the density estimated using the method above. It shows that the density estimator used here can distinguish between low-density void regions and high-density group or cluster environments.

\subsection{Neighbor selection} \label{sec:nn}

We take galaxies in sample $S_{\rm all}$ for the finding of the neighbors. In order for a galaxy to be classified as a neighbor to an AGN, we use the following conditions:

\begin{itemize}
	\item Stellar mass of the neighbor should be more than half of AGN, i.e., $M^{}_{\rm *,neigh}/M^{}_{\rm *,AGN}>0.5$. This choice is motivated by a major merger scenario in which a neighbour with comparable mass will have significant effect.
	\item The galaxy which has a minimum $d_{\rm neigh}/R_{\rm vir,neigh}$ ratio is taken, where $d_{\rm neigh}$ is the distance to the galaxy and $R_{\rm vir,neigh}$ is the virial radius of the galaxy marked as the neighbor. 
\end{itemize}

We note here that we consider only the effect of the nearest neighbor. In the literature, a distance cut is used within which all the neighbors are used for analysis, and a tidal force estimator (which depends on the ratio of mass, the radius of AGN host galaxy, and distance to the neighbor) is used to study the environmental effect \citep{2013Sabater,2015Sabater}. Our method does not use such distance cuts for neighbor identification. Instead, the dependence on the mass of the neighbor comes through the virial radius of the neighbor \citep{2007Park,2009Park,2009Hwang}. As will be discussed in Figure \ref{fig:envfrac}, the fraction of AGNs does not change with the background density. This supports our simple assumption that the nearest neighbor can capture the environmental effect of the most influencing neighbor rather than the collective effect of galaxies beyond the nearest neighbor.

\section{Results}\label{sec:results} 

In the following subsections, we will discuss the results of our analysis. In Section \ref{sec:props}, we explore the effects of the intrinsic properties of AGN host galaxies on the bolometric luminosity of AGNs. We then study the impact of the environment in Section \ref{sec:local}.

\subsection{Intrinsic properties}\label{sec:props} 

The intrinsic properties we focus on in this study are specific star formation rate (sSFR), gas metallicity ($Z_{\rm gas}/Z_{\rm NonAGN}$), morphology ($A/A_{\rm NonAGN}$), and kinematics ($(v_{\rm rot}/\sigma)/(v_{\rm rot}/\sigma)_{\rm NonAGN}$) of the host galaxy. As mentioned above all these properties have been scaled with the stellar mass-matched values from the Non-AGN galaxies.

\begin{figure} 
	
	\centering
	
	\includegraphics[width=0.45\textwidth]{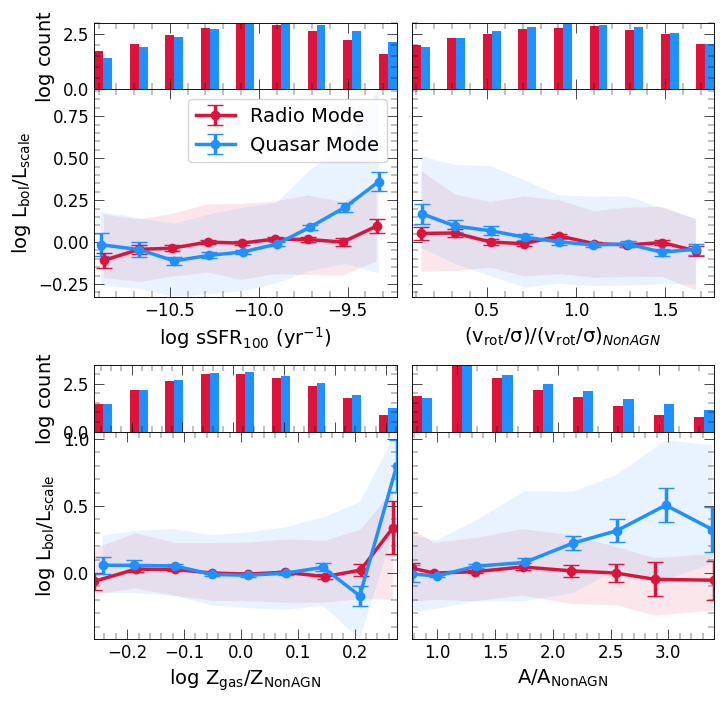}
	
	\caption{Variation of bolometric luminosity of AGN galaxies ($L_{\rm bol}$) with specific star formation rate averaged over 100 Myr timescale ($\rm sSFR_{\rm 100}$;top-left),($(v_{\rm rot}/\sigma)/(v_{\rm rot}/\sigma)_{\rm NonAGN}$; top-right), asymmetry parameter ($\mathrm{A/A_{NonAGN}}$; bottom-right), and metallicity of gas ($Z_{\rm gas}/Z_{\rm NonAGN}/Z_{\rm NonAGN}$;bottom-left) in the subsample $S_{\rm bin}$. The two colors represent the two modes of AGN, quasar (blue) and radio (red). The shaded region associated with each curve depicts the 0.25 and 0.75 quartile ranges. The histogram on top shows the number of AGN galaxies in each bin for the two modes. Both modes show an increase in the bolometric luminosity with $\rm sSFR_{\rm 100}$. Quasar mode shows an increase of bolometric luminosity with an increase in $Z_{\rm gas}/Z_{\rm NonAGN}$. The high metallicity gas can cool giving rise to an increase in bolometric luminosity if the origin of AGN activity in the quasar is intrinsic. A mild increase of $< 0.25$ dex in quasar mode with the decrease in $(v_{\rm rot}/\sigma)/(v_{\rm rot}/\sigma)_{\rm NonAGN}$ is found, and bolometric luminosity of the quasar mode AGN increases with increasing the anisotropic parameter ($A/A_{\rm NonAGN}$).} \label{fig:LvsSFR} 
\end{figure}

\begin{figure} 
	
	\centering
	
	\includegraphics[width=0.45\textwidth]{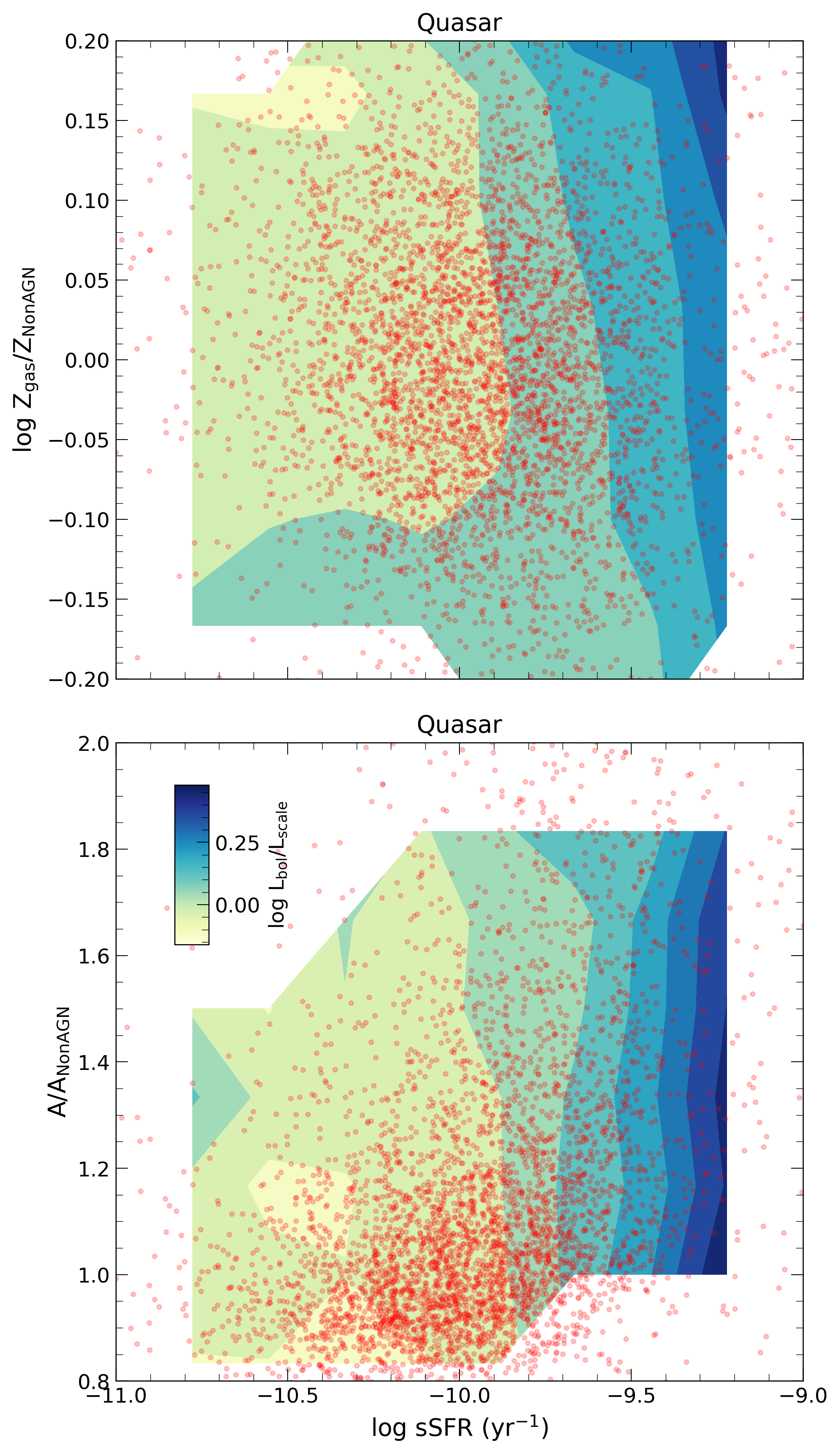}
	
	\caption{Contours of bolometric luminosity in quasar mode in sSFR-$A/A_{\rm NonAGN}$ plane and sSFR-$Z_{\rm gas}/Z_{\rm NonAGN}$ plane. The red points show the distribution of quasars in the two planes. For a given value of bolometric luminosity, the contours are aligned vertically. Therefore, the specific star formation rate (sSFR) is the dominant intrinsic property determining the bolometric luminosity of the AGN in quasar mode. Note that for low sSFR ($<10^{-10} \ yr^{-1}$)and low gas metallicity, the contours are not vertical and bolometric luminosity changes with the decrease in gas phase metallicity.}
	\label{fig:cont_int} 
	
\end{figure}

\textit{Specific star formation activity}: If the AGN activity of the galaxy is fuelled by the presence of gas in the galaxy, then it is expected that this gas present would also lead to an increase in the star formation activity in the galaxy. \cite{2009Choi} using SDSS reported that late-type galaxies are the dominant hosts of AGNs, and Eddington line ratio luminosity is higher for blue galaxies. \cite{2012Mullaney} performed a X-ray stacking analysis for star forming galaxies at $z\sim1$ and $z\sim2$.  They reported no correlation between star formation rate (SFR) and AGN activity. 

On the other hand, for the AGNs with high luminosity \cite{2008Lutz} used mid-infrared spectroscopy from $Spitzer$ for 12 galaxies and found a correlation between AGN luminosity and star formation indicators PAH and FIR continuum. \cite{2012Rosario} using deep far-infrared (FIR) and X-ray data from $Herschel$ and $Chandra$ respectively found that a high star formation rate is correlated with high AGN activity at $z<1$. \cite{2014Hickox} used the time variability of AGNs and concluded that SFR (measured using FIR) is weakly correlated with AGN luminosity in the redshift range $0<z<2$. \par 

In Figure \ref{fig:LvsSFR} we show the correlation between specific star formation rate averaged over 100 Myr ($\rm sSFR_{100}$) and the AGN bolometric luminosity ($L_{\rm bol}/L_{\rm scale}$) in the subsample $S_{\rm bin}$. For both the modes, radio (red) and quasar (blue), the AGN bolometric luminosity increases with an increase in $\rm sSFR_{\rm 100}$. The median increase in quasar mode is more significant (0.5 dex) than in radio mode ($\le 0.25$ dex). This indicates that the AGN bolometric luminosity is correlated to the star formation activity of the galaxy. We observe a half order of magnitude of change in the bolometric luminosity with $\rm sSFR_{\rm 100}$ in the quasar mode. \par 

Our results compare well with \cite{2015Sijacki}. They used the Illustris simulation to study the correlation between mock AGN hard X-ray luminosity with the star formation rates within the stellar half-mass radius. They also found a weak correlation with a large scatter, similar to our result. They compared their results with \cite{2015Azadi} for different redshifts and found the weak scatter and trend to be present up to $z\sim1$. The large scatter they concluded can be attributed to physical processes that act on time scales shorter than that of star formation \citep{2014Hickox}. Recently, \cite{2022Mountrichas}, who studied the star formation rate of galaxies in the COSMOS-Legacy Survey in the redshift range $0<z<2.4$. For the X-ray luminosities $L_{\rm X,2-10keV} > 2-3 \times 10^{44} \rm erg \ s^{-1}$ there is an increase of star formation rate of AGN compared to other star-forming galaxies. This result matches our finding of specific star formation being higher for high AGN luminosity. \par

\textit{Metallicity}: The metallicity of the stars and gas in a galaxy is related to its cosmic history. A galaxy evolves its metallicity by recycling material via intrinsic star formation or via accreting gas from its environment. Gas metallicity in galaxies is weakly correlated with the galaxy luminosity by well-known metallicity-luminosity relation \citep{1993Hamann,2002Shemmer}. In Figure \ref{fig:LvsSFR} we show the variation of bolometric luminosity of AGNs with gas metallicity in the subsample $S_{\rm bin}$. Quasar mode shows an increase in the bolometric luminosity with $Z_{\rm gas}/Z_{\rm NonAGN}$ of $\simeq 0.5$ dex, whereas no change is observed for radio mode galaxies. 
If the origin of AGN activity is intrinsic (i.e., accretion from gas in the host galaxy) high metallicity gas can cool giving rise to an increase in bolometric luminosity. We explore the origin of AGN activity in later sections. The results indicate that quasar mode AGNs with more processed gas are more luminous. We observe a 0.25 dex increase in AGN activity for low metallicity but because of the limitation on the number of AGNs, it requires more investigation.

\textit{Kinematics and morphology}: In the hierarchical model of galaxy formation, galaxies assemble mass by mergers throughout their evolution. These events significantly affect the kinematics and morphology of the galaxy. In the right panels of Figure \ref{fig:LvsSFR}, we show the relation between bolometric luminosity and kinematic properties of AGN host galaxies. The bolometric luminosity in quasar mode shows a mild increase of $<0.25$ dex with the change in $(v_{\rm rot}/\sigma)/(v_{\rm rot}/\sigma)_{\rm NonAGN}$. Here $v_{\rm rot}$ and  $\sigma_{\rm d}$ are the mass-weighted rotational velocity and velocity dispersion respectively. However, bolometric luminosity increases with the increase in an asymmetry of an AGN-host galaxy for quasar mode with a median change of $\simeq$ 0.5 dex.

We find that bolometric luminosity in quasar mode is significantly correlated with specific star formation. On the other hand, the specific star formation rate, gas metallicity, and asymmetry of the host galaxy could have mutual correlations with the star formation. In order to study which one is the dominant property among them, in Figure \ref{fig:cont_int} we plot the contours of bolometric luminosity in the $\rm sSFR_{\rm 100}$-$Z_{\rm gas}/Z_{\rm NonAGN}$ and $\rm SFR_{\rm 100}$-$A/A_{\rm NonAGN}$ planes.  For a given value of bolometric luminosity, the contours are aligned vertically. As expected, the bolometric luminosity steadily increases from a low to a high specific star formation rate. Therefore, we conclude that the specific star formation rate of the host galaxy is the dominant intrinsic property correlated with the AGN activity, whereas bolometric luminosity increases weakly with gas metallicity and has no net dependence on the asymmetry of galaxies. Note that for low sSFR ($<10^{-10} \ yr^{-1}$) and low gas metallicity the contours are not vertical and bolometric luminosity changes with the decrease in gas phase metallicity. Due to less number of galaxies in this region, this requires further investigation in future work.

\subsection{Effects of local environment}\label{sec:local} 

We note that most AGNs of HR5 galaxies have bolometric luminosity lower than $\rm 10^{45} erg \ s^{-1}$, which is below the quasar definition widely used in observations: B-band absolute magnitude greater than -23 mag ($\approx L_{\rm bol}\gtrsim 10^{45} \rm erg \ s^{-1}$). Therefore, the AGNs in HR5 are not expected to instantaneously make an observable impact on their hosts or the surrounding environment \footnote{We note here that interpreting the weaker trends (minor change with respect to a scatter) requires a caution that the observed trend could be a representation of a new secondary correlation.}. The top left panel of Figure~\ref{fig:LvsSFR}  shows that HR5 quasars do not quench the SFR much instantaneously, complementing the results of \cite{2012Hopkins} who reported that the delay in SFR and AGN activity could arise because of dynamical effects. Recently, \cite{2021Jun} studying the role of radiation pressure in blowing out obscured quasar observed the timescale of AGN activity to be smaller than the outflows. At $z=0.625$, the radio mode feedback will not have a strong effect on the large-scale environment seen at low redshift  \citep[e.g.][]{2012Fabian}. Hence, our study explores the environment's effect on AGN activity rather than vice-versa. \par

We study the effects of the local environment in the following subsections. First, we begin our discussion by examining the abundance of AGNs in different background densities in Section \ref{sec:frac}. Then we study the effects of the environment on different properties in Section \ref{sec:localenv}.

\subsubsection{Fraction of AGNs}\label{sec:frac} 

\begin{figure}
	
	\centering
	
	\includegraphics[width=0.45\textwidth]{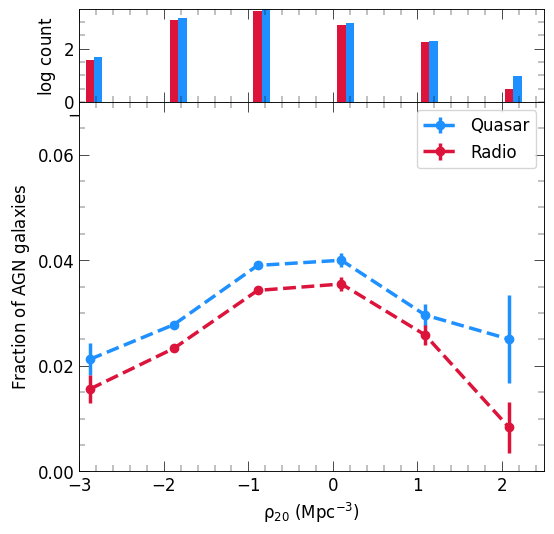}
	
	\caption{Fraction of AGN galaxies in different bins of background density $\rho_{20}$.  The fraction of AGNs increases till intermediate background densities and decreases afterward. The fraction of radio and quasar mode galaxies are similar to each other for a range of background densities. Environmental effects can lead to a reduction in the gas supply (circumgalactic medium; CGM) believed to fuel the radio-mode AGN, resulting in a lower number of radio-mode galaxies in high-density environments. The fraction of AGNs remains only between $1-5$ \% for each mode. }
	
	\label{fig:envfrac}
	
\end{figure}

In Figure \ref{fig:envfrac} we show the fraction of galaxies hosting AGN compared to all the galaxies in different background density bins for the sample $S_{\rm all}$. The fraction of AGNs increases till intermediate background densities and decreases afterward. The fraction of radio and quasar mode galaxies are similar to each other for a range of background densities. At all the background densities the fraction of AGNs remains only between $1-5$ \% for each mode. \par 

At the highest density bin, a fraction of AGNs in quasar mode is higher than in the radio mode. Previous works have shown the cold gas accretion to be associated with quasar mode and hot gas accretion to power the radio mode AGNs (for example \cite{2016Argudo}). The hot gas in galaxies is stripped due to environmental effects in high-density environments. This may lead to less number of radio mode AGNs to be found in high-density environments. We will explore this scenario of AGN triggering in the following sections. This conclusion agrees with \cite{2020Mishra} using X-ray data from $ROSAT$ at $z<0.5$ showed that the fraction of AGNs in clusters ($\sim 5$\%) is lower than the field ($\sim 1$\%). They concluded that the relatively lower fraction is due to a dearth of cold gas due to environmental effects. \cite{2021Duplancic} for galaxies in redshift range ($0.05 \leq z \leq 0.15$) used data from SDSS and Wide-field Infrared Survey Explorer (WISE) to report a higher fraction of AGNs at intermediate densities (pairs and triplets) than the corresponding fraction in groups. Our results agree well with their finding. \par 

The low variation in the fraction of galaxies with the environment has also been reported at low redshifts ($z<0.1$). \cite{2001Carter},  \cite{2003Miller}, and \cite{2019Amiri}, reported that the fraction of galaxies with an AGN remains similar from high-density clusters to the low-density field environment. At similar redshift, \cite{2019Man} defined AGN fraction as AGNs relative to star-forming galaxies and found its dependence on overdensity, central/satellite, and group halo mass to be very little.

\subsubsection{Variation of AGN properties with local environment} \label{sec:localenv} 

In this subsection, we study how the local environment affects the properties of AGNs in $S_{\rm bin}$. We use the local number density around an AGN ($\rho^{}_{\rm 20}$) and distance to the neighbor normalized by its virial radius ($d_{\rm neigh}/R_{\rm vir,neigh}$) as the proxy for the effects of the environment. This comparative analysis will help us understand if the change in properties of the AGNs in both modes is driven by the cumulative effect of the number of galaxies present in the neighborhood or by one-to-one interaction with the most influential neighbor. \par

\begin{figure} 
	
	\centering
	
	\includegraphics[width=0.45\textwidth]{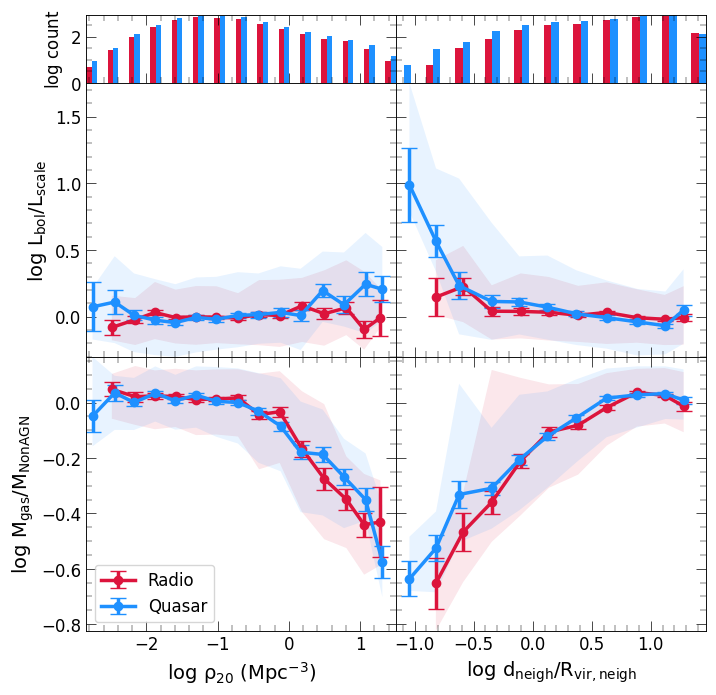}

	\caption{Variation of bolometric luminosity of AGN galaxies ($L_{\rm bol}/L_{\rm scale}$; top) and gas mass ($\frac{M_{gas}}{M_{NonAGN}}$; bottom) with background density and distance to the neighbor in the subsample $S_{\rm bin}$. The rest of the details are the same as Figure \ref{fig:LvsSFR}. We observe a mild rise of $<0.25$ dex in quasar mode with background density. There is a significant ($\sim 1$ dex) increase in bolometric luminosity very close to the one-third virial radius of a neighboring galaxy for quasar mode. Radio mode shows no dependence on an increase in background density or distance to the neighbor. The gas mass in the host galaxy decreases for both modes in higher background density and smaller distances to the nearest neighbor. }
	
	\label{fig:fvsL}
	
\end{figure}

\textit{Bolometric Luminosity}: In Figure \ref{fig:fvsL} (top) we show the effect on the bolometric luminosity ($L_{\rm bol}/L_{\rm scale}$) of AGNs by the local environment. We note that the quasar mode is significantly affected by the environment. There is a rise in bolometric luminosity in the quasar mode from low-density void to high background density groups or cluster environments of $\leq 0.25$ dex. The radio mode shows no significant increase. We also observe a steep increase in the bolometric luminosity of $\sim$1 dex in the quasar mode if the host galaxy of AGN is very close to the one-third virial radius of a neighboring galaxy. The effect on the bolometric luminosity is negligible if the host galaxy is more than one one-third virial radius away from the neighbor. Radio mode, on the other hand, is unaffected by the presence of a nearby galaxy with variations $\leq 0.1$ dex .\par

\begin{figure} 
	
	\centering
	
	\includegraphics[width=0.45\textwidth]{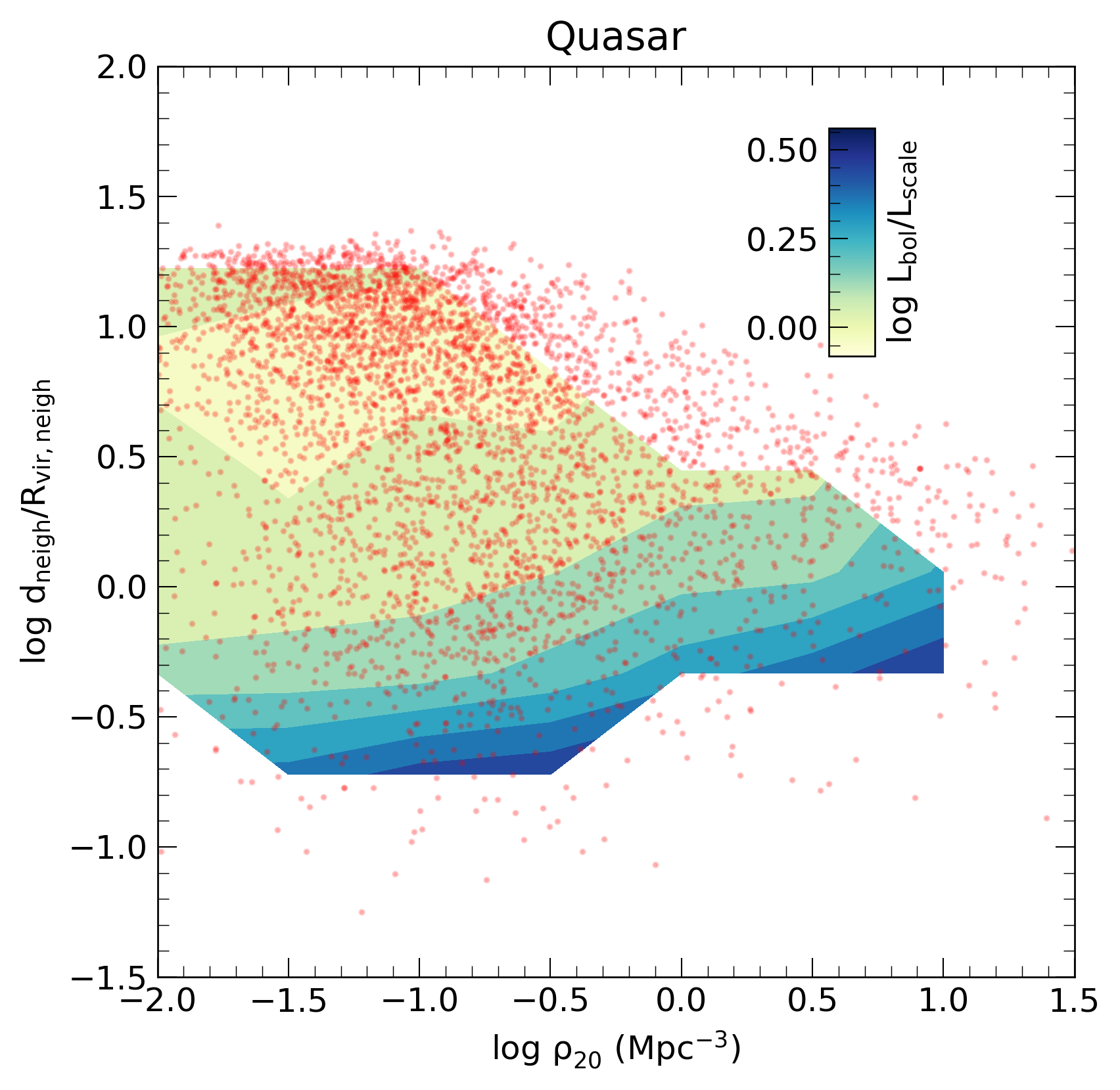}
	
	\caption{Contours of bolometric luminosity in quasar mode in  $\rho^{}_{20}-d_{\rm neigh}/R_{\rm vir,neigh}$ plane. The red points show the distribution of quasars in the plane. For a given background density, as the distance to the nearest neighbor is decreased, the bolometric luminosity increases. When $log \ d_{\rm neigh}/R_{\rm vir,neigh} \leq 0$, $L_{\rm bol}/L_{\rm scale}$ is a monotonically increasing function of $d_{\rm neigh}/R_{\rm vir,neigh}$.  The contours are slanted horizontally indicating the distance to the neighbor is the dominating parameter. The non-horizontal features toward higher local density could be some additional environmental effects
	from the 2nd and 3rd(etc.) nearest neighbors as they are getting closer. } \label{fig:cont_lbol} 
	
\end{figure}

There is a degeneracy between the effect of background density and neighbor on the bolometric luminosity of quasar mode AGNs. To compare these two competing effects, in Figure \ref{fig:cont_lbol} we show the contours of the bolometric luminosity in $\rho^{}_{20}-d_{\rm neigh}/R_{\rm vir,neigh}$ plane. The contours are slanted horizontally indicating the distance to the neighbor is the dominating parameter. The low values of distance to neighbor correspond to galaxies having close interaction with their neighbor. The AGNs that have close neighbors ($log d_{neigh}/R_{vir,neigh}<0$) also reside in high-density regions. We note that the contours show some non-horizontal features toward higher local density. It could be some additional environmental effects
from the 2nd and 3rd(etc.) nearest neighbors as we go to high-density regions as they are getting closer.

\begin{figure} 
	
	\centering
	
	\includegraphics[width=0.45\textwidth]{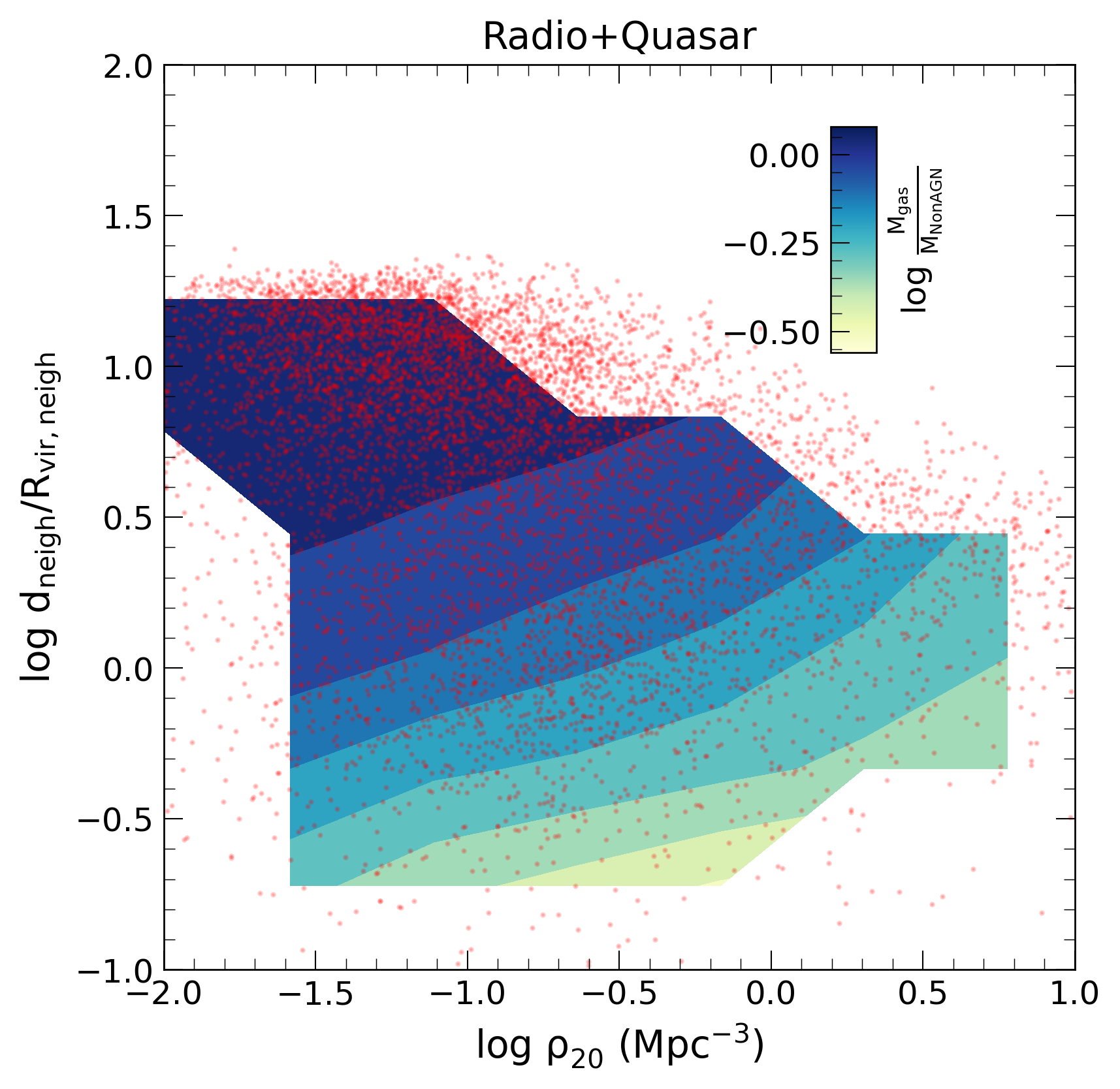}
	
	\caption{Contours of total gas mass ($\frac{M_{gas}}{M_{NonAGN}}$) in AGN host galaxies in $\rho^{}_{20}-d_{\rm neigh}/R_{\rm vir,neigh}$ plane. The red points show the distribution of AGNs in the plane. Gas mass in the host galaxy of AGN decreases in high background density as the distance to the nearest neighbor decreases. The lack of vertical or horizontal alignment in the contours suggests that AGN host galaxies in the local high-density environment or undergoing interaction exhaust their gas.} \label{fig:cont_gas} 
	
\end{figure}

\textit{Mass of gas}: In Figure \ref{fig:fvsL} (bottom), we plot the variation of the gas of the host galaxy with the parameters of the local environment. It shows that the gas mass in the host galaxy decreases for both modes in high background density by $\sim$0.4 dex compared to low background density. If the host galaxy is within one virial radius of a neighboring galaxy, the gas in the host galaxy also decreases significantly by $\sim$0.8 dex compared to the AGNs beyond one virial radius of the neighbor. In order to compare the effects of the two local parameters in Figure \ref{fig:cont_gas} we show the contours of $M_{\rm gas}/M_{NonAGN}$ in  $\rho^{}_{20}-d_{\rm neigh}/R_{\rm vir,neigh}$ plane. The contours are not either horizontal or vertical for both radio and quasar mode AGNs. This indicates that both high local density and interactions with near neighbors make AGN host galaxies exhaust their gas.\par

\begin{figure}
	
	\centering
	
	\includegraphics[width=0.45\textwidth]{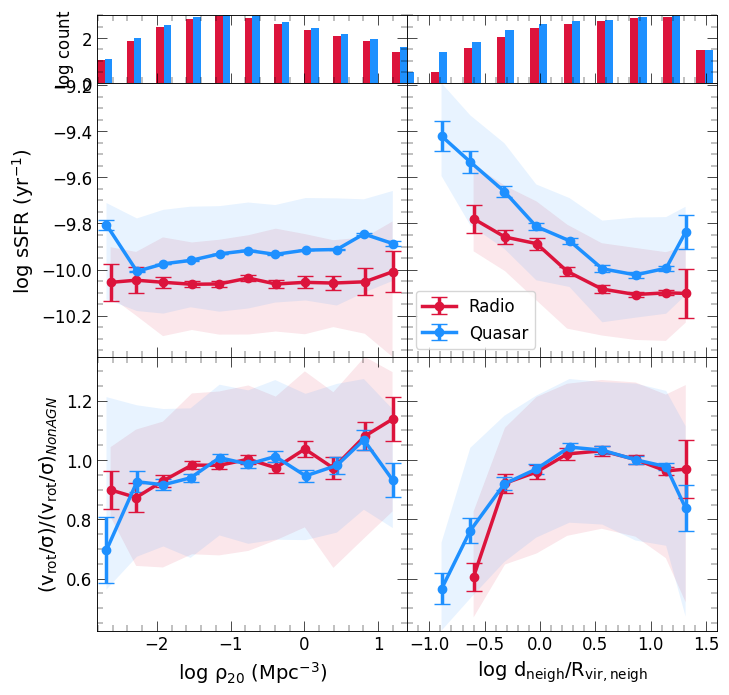}
	
	\caption{The specific star formation rate (top) in both modes shows an increase with background density. The rest of the details are the same as Figure \ref{fig:LvsSFR}. SFR in quasar mode increases significantly when the distance from the nearest neighbor decreases. The kinematic parameter ($(v_{\rm rot}/\sigma)/(v_{\rm rot}/\sigma)_{\rm NonAGN}$; bottom) increases with background density. It shows a significant decrease with a decrease in distance to the nearest neighbor.}
	
	\label{fig:fvsSFR}
	
\end{figure}

\textit{Specific star formation rate}: Star formation activity of a galaxy is influenced by its environment. In Figure \ref{fig:fvsSFR} (top), we show the variation of specific star formation rate averaged over 100 Myr ($ \rm sSFR_{\rm 100}$) with background
density (left) and distance to neighbor (right) in the subsample
$S_{\rm bin}$. The specific star formation activity in quasar modes increases with background density by 0.2 dex.The trend is evident but very weak, as the increase is within the scatter. In both modes, the sSFR increases significantly when the distance from the neighbor decreases. In the lowest local density bin and highest distance to neighbor bins there is an increase which can be due to recent mergers but we leave the further analysis for future as number of galaxies are less.\par 

\begin{figure} 
	
	\centering
	
	\includegraphics[width=0.45\textwidth]{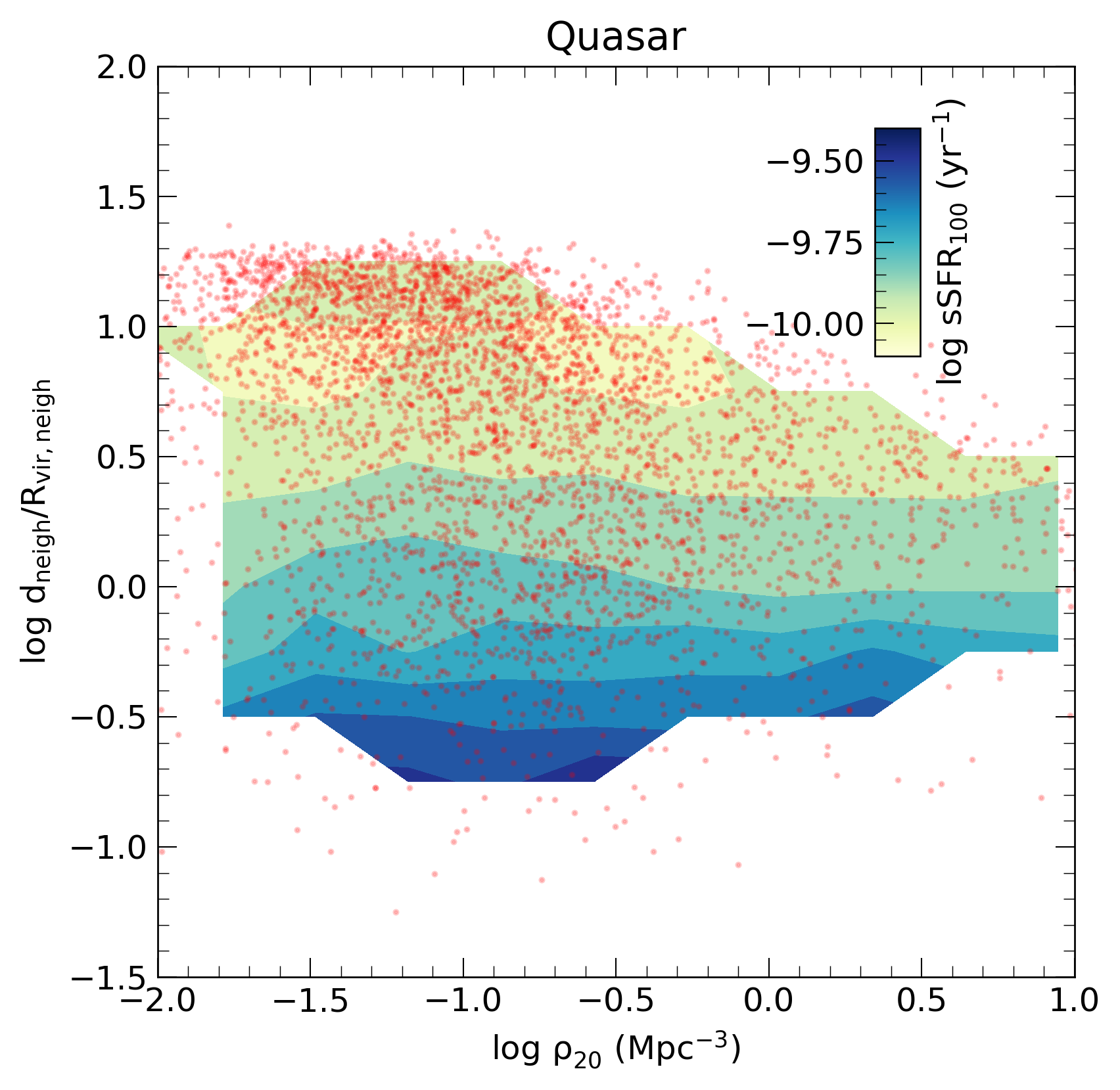}
	
	\caption{Contours of specific star formation rate ($\rm sSFR_{\rm 100}$) in quasar mode in  $\rho^{}_{20}-d_{\rm neigh}/R_{\rm vir,neigh}$ plane. The red points show the distribution of quasars in the plane. For a given $\rm SFR_{\rm 100}$ value, contours are aligned horizontally, indicating that $d_{\rm neigh}/R_{\rm vir,neigh}$  is the dominant quantity in determining SFR in presence of a close neighbor. The high values of sSFR at low neighbors correspond to currently interacting and recent major merger cases respectively.} \label{fig:cont_sfr} 
	
\end{figure}

To explore the degeneracy in the quasar mode in Figure \ref{fig:cont_sfr} we plot the contours of star formation rate ($\mathrm{sSFR_{100}}$) in the  $\rho^{}_{20}-d_{\rm neigh}/R_{\rm vir,neigh}$ plane for quasar mode. The contours are horizontal indicating that $d_{\rm neigh}/R_{\rm vir,neigh}$  is the dominant quantity in determining sSFR in presence of a close neighbor. The high values of sSFR at the low distances to the neighbor correspond to currently interacting major merger cases \citep{2009ParkChoi}. \par

\textit{Kinematics}: We characterize the kinematics properties of the AGN host galaxies by the parameter ($(v_{\rm rot}/\sigma)/(v_{\rm rot}/\sigma)_{\rm NonAGN}$).  In the bottom panels of Figure \ref{fig:fvsSFR}, we show the variation of the kinematics parameter with the background density and the distance to the neighbor ($d_{\rm neigh}/R_{\rm vir,neigh}$). It is evident that for both modes, the kinematics of the AGN host galaxies is affected by both the background density and distance to the neighbor. The AGN host galaxies become rotation dominated in the regions with high background density, but the trend is very weak and within the scatter. If the AGN host galaxy is within the virial radius of a neighboring galaxy, the discs become pressure-supported. The environment parameters, background density, and distance to neighbor thus have the opposite effect. \par 

\begin{figure} 
	
	\centering
	
	\includegraphics[width=0.45\textwidth]{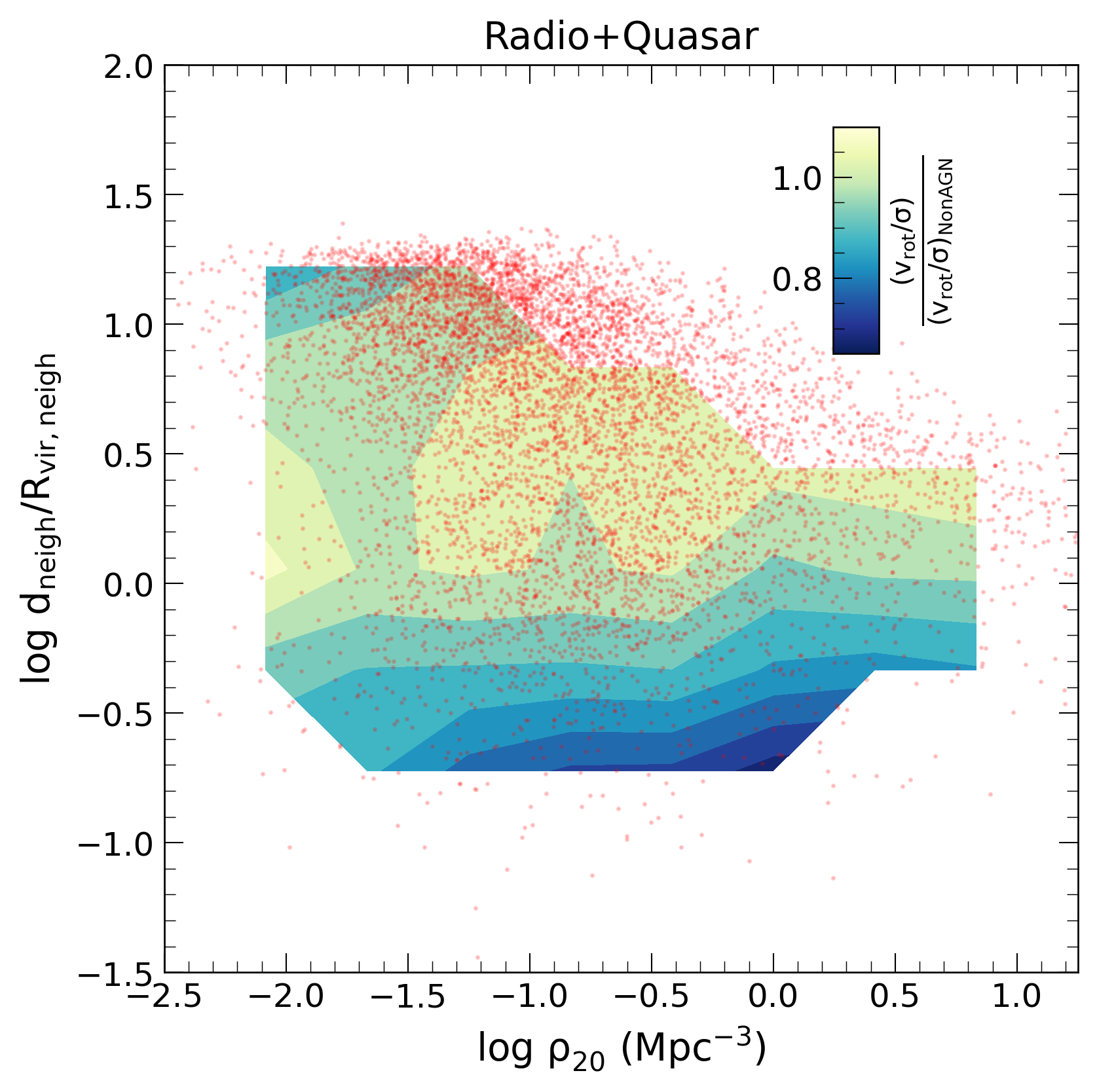}
	
	\caption{Contours of kinematics of host galaxy of AGNs characterised by $(v_{\rm rot}/\sigma)/(v_{\rm rot}/\sigma)_{\rm NonAGN}$  in  $\rho^{}_{20}-d_{\rm neigh}/R_{\rm vir,neigh}$ plane. 
	The red points show the distribution of AGNs in the plane.  For a given $(v_{\rm rot}/\sigma)/(v_{\rm rot}/\sigma)_{\rm NonAGN}$, contours are aligned horizontally for two extreme values of $log \ d_{neigh}/R_{vir,neigh}$. The AGNs in the region with  $log \ \rho^{}_{20}<-1$ and $log \ d_{\rm neigh}/R_{\rm vir,neigh}>0$ are isolated with no or low merger rate, therefore, have rotation dominated kinematics. Host galaxies with close neighbors have pressure-supported kinematics. For the distance $0<log \ d_{\rm neigh}/R_{\rm vir,neigh}<1$ the contours are not horizontal indicating the effect of the neighbor on host galaxy kinematics of AGN is not dominant if the distance to the neighbor is more than one virial radius.} \label{fig:cont_vrot} 
	
\end{figure}

In Figure \ref{fig:cont_vrot} we show the plot of contour of $(v_{\rm rot}/\sigma)/(v_{\rm rot}/\sigma)_{\rm NonAGN}$ in the $\rho^{}_{20}-d_{\rm neigh}/R_{\rm vir,neigh}$ plane for both the modes. For a given $v_{\rm rot}/\sigma_{\rm d}$, contours are aligned horizontally for two extreme values of $log \ d_{neigh}/R_{vir,neigh}$. The AGNs in the region with  $log \ \rho^{}_{20}<-1$ and $log \ d_{\rm neigh}/R_{\rm vir,neigh}>0$ are isolated with no or low merger rate \citep[figure 3 of ][]{2009ParkChoi}. Therefore, they are expected to have rotation-dominated kinematics. AGN host galaxies with close neighbors have pressure-supported kinematics. Therefore, distance to the neighbor is the dominant parameter compared to background density. This indicates that the presence of the neighbor ($d_{\rm neigh}/R_{\rm vir,neigh}$) regulates the kinematics of the AGN host galaxy in all the local environments. For the distance $0<log \ d_{\rm neigh}/R_{\rm vir,neigh}<1$ the contours are not horizontal indicating the influence of neighbor fades off if the distance is more than one virial radius.

\section{Discussion}\label{sec:diss}

The large volume of HR5 enables us to explore different environment scales while maintaining a resolution of 1 kpc. The large box size also helps explore the AGNs in two modes in different stages of evolution. In studying the effect of the environment of AGNs, we are thus able to maintain a good statistical sample size in each bin even after distinguishing the two modes and removing the effects of secular evolution. We also note that thermal efficiency is about 10 times larger in kpc-scale resolution cosmological simulations compared to observations or isolated galaxy simulations \citep[for example;][]{2011Wang}. This might yield biased quasar bolometric luminosities and number densities which in turn can bias the comparison between simulations and observations. Since our study only compares the luminosity/accretion rate trend rather than absolute numbers, this should not alter the results. In the following text, we compare our results with the existing literature.

In Section \ref{sec:props}, we observed that the star formation rate is the fundamental property that is linked to the AGN activity. We showed in Section \ref{sec:local} that distance to the neighbor is the dominant property of the environment. The gas mass is less, and AGN activity is stronger closer to the neighbor. These results indicate that the gas supply inside the galaxy is fuelling the AGN activity. In this section, we explore the origin of this gas in the host galaxy, i.e., if it is intrinsic in nature or acquired from the environment. \par

\begin{figure} 
	
	\centering
	
	\includegraphics[width=0.45\textwidth]{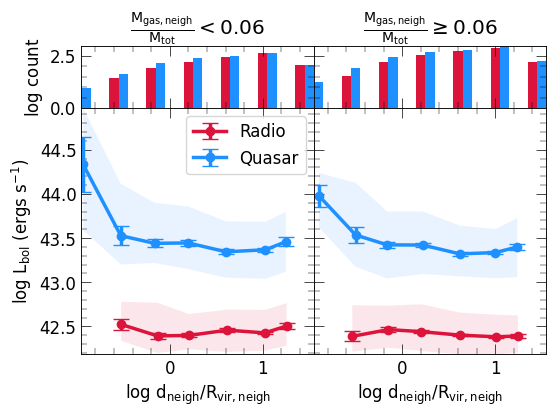}
	
	\caption{Bolometric luminosity ($L_{\rm bol}$) of AGNs with distance to the neighbor for gas-rich and gas-poor neighbors. The rest of the details are the same as Figure \ref{fig:LvsSFR}. The increase in $L_{\rm bol}$ is independent of the gas fraction in the neighbor. This indicates that the gas in the host galaxy of AGNs is not likely coming from the neighbor. } \label{fig:neigh_gas} 
	
\end{figure}

In Figure \ref{fig:neigh_gas} we divide the neighbors of the AGNs into gas-rich ($\frac{M^{}_{\rm gas,neigh}}{M^{}_{\rm tot}} \geq 0.06$) and gas-poor ($\frac{M^{}_{\rm gas,neigh}}{M^{}_{\rm tot}} < 0.06$) where $M_{\rm tot}$ is total mass of the subhalo. The value of 6 \% is chosen to maintain a similar population in both bins of the gas fraction. It is evident that the trend of increase in bolometric luminosity is independent of the gas mass fraction of the neighbor. We conclude that the gas fuelling the AGN activity is not accreted from a near neighbor. As we show in Figure \ref{fig:envfrac} that the fraction of AGNs does not change with the background environment density, we conclude that the gas fuelling in AGN activity has an intrinsic origin. \par

Assuming that AGNs are fueled by the accretion of cold gas in the vicinity (circumgalactic medium; CGM) of the galactic center. Since quasar mode has higher luminosity (higher accretion) the amount of cold gas present near the galactic center of the galaxy in quasar mode should be more than a galaxy in radio mode. has higher AGN activity and gas fuelling the AGN activity has an intrinsic origin, the host of quasar mode galaxies should have higher cold gas near the nuclear region.

\begin{figure} 
	
	\centering
	
	\includegraphics[width=0.45\textwidth]{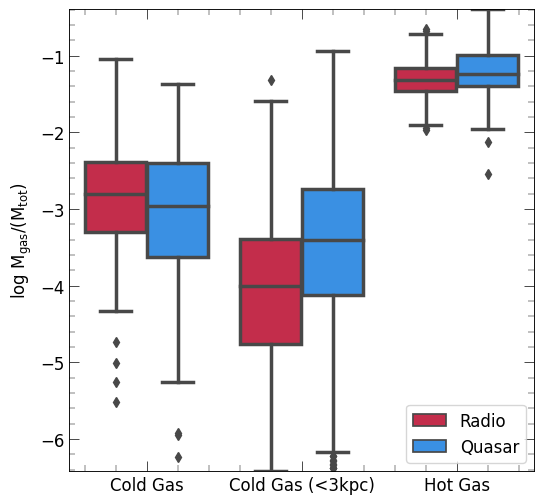}
	
	\caption{Total cold gas mass, the cold gas mass inside 3 ckpc from center, and total hot gas mass of the host galaxies in two modes. The horizontal lines in filled regions show the first quartile (Q1), median and third quartile (Q3) in increasing order. Points represent values beyond the minimum (Q1 - 1.5 $\times$ IQR)  and maximum (Q3 + 1.5 $\times$ IQR) represented by horizontal lines in increasing order, where IQR is the interquartile range. The cold mass content in the quasar mode is more than in the radio mode, especially in the inner regions of the host galaxy.} \label{fig:comb_AGN} 
	
\end{figure}

 In Figure \ref{fig:comb_AGN}, we show the gas present in the quasar and radio mode AGNs. The hot gas content ($T> \rm 10^4 \ K$) and cold gas ($T \leq \rm 10^4 \ K$) of the AGN host galaxies are similar in both modes. The contrast becomes more striking when the gas within the inner 3 kpc of the center of galaxies is considered. Therefore, we conclude that the gas in the AGN activity is related to the cold gas within the vicinity (circumgalactic medium; CGM) of the nuclear region. The hot gas ($T > 10^4 K$) supply in the radio mode AGNs is higher than in the quasar mode. Figure \ref{fig:envfrac} shows that the fraction of radio mode AGNs in high density decreases. In Figure \ref{fig:fvsL} we observed low effect of interaction on radio mode AGNs. These results point to hot gas accretion from the vicinity (circumgalactic medium; CGM) as the fuelling mechanism of the radio mode AGNs.\par

\begin{figure} 
	
	\centering
	
	\includegraphics[width=0.45\textwidth]{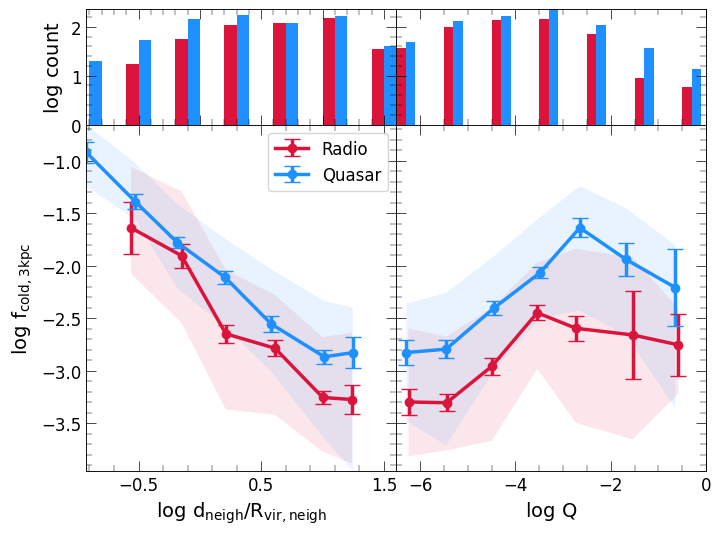}
	
	\caption{Cold gas ($T< \rm 10^4 \ K$) fraction ($f_{g}=M_{cold,gas}/M_{tot,gas}$) inside 3 ckpc from center AGNs with distance from the neighbor (left) and tidal force estimator ($Q$; right). The rest of the details are the same as Figure \ref{fig:LvsSFR}. With a decrease in the distance to the neighbor and an increase in the tidal interaction with neighbors, gas in the host galaxy is pushed toward the central regions of the galaxy.  } \label{fig:innerq} 
	
\end{figure}

In order to test if the cold gas inside the galaxy is being pushed to the center of the galaxy fueling the AGNs in quasar mode in Figure \ref{fig:innerq} (left) we show the effect of the distance to the neighbor on the cold gas fraction ($=M_{cold,gas}/M_{tot,gas}$) in the inner 3 kpc ($f_{cold,3kpc}$) in the host galaxies for the two modes. It can be seen that for both modes, the cold gas fraction within the inner 3 kpc of the host galaxy increases by $> 1$ dex as the distance to the neighbor decreases. Thus, the presence of a neighbor helps decrease the angular momentum of the gas due to dynamical friction and tidal effects to funnel it toward the center \citep{2006Donghia} of the galaxy in quasar mode more efficiently, leading to higher AGN activity. \par 

We consider a cutoff distance of 3 cMpc from the AGN and use the tidal estimator, which compares the tidal forces with the internal binding force of the galaxy \citep{2007Verley,2013Sabater} given by:

\begin{eqnarray}
	Q_{\rm t} \equiv \log \left( \sum _{i} \frac{M_{\rm AGN}}{M_{i}} \left( \frac{2R_{\rm AGN}}{d_{i,t}} \right)^{-3} \right),
\end{eqnarray}

where $M^{}_{\rm AGN}$ and  $M^{}_{\rm i}$ is the total mass of the galaxy and $i$-th neighbor respectively,  $R^{}_{\rm AGN}$ is the half stellar mass radius of the galaxy and $d_{i,t}$ is the distance to the $i$-th neighbor. In Figure \ref{fig:innerq} (right) we show the cold gas fraction inside the inner 3 ckpc of the galaxy with a tidal estimator. It is evident that with an increase in the strength of the tidal forces, the gas in the inner regions of the AGN increases by $> 0.5$ dex for both modes, but radio mode shows a large scatter. Therefore, we conclude that the tidal interaction with the neighbors pushes the gas inside the galaxy toward the central region, increasing the AGN activity.\par

\begin{figure} 
	
	\centering
	
	\includegraphics[width=0.45\textwidth]{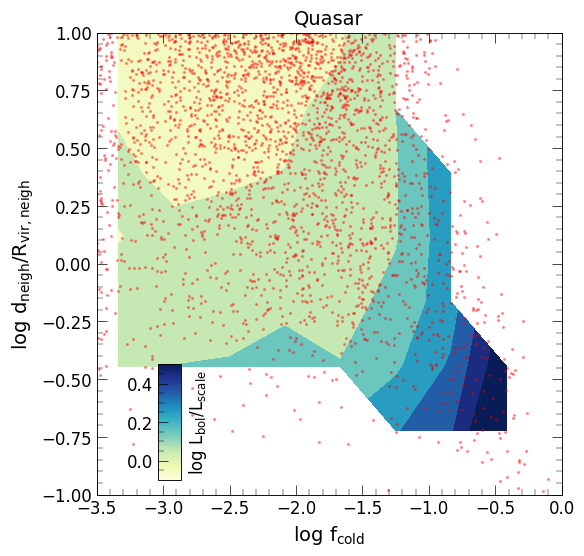}
	
	\caption{Contours of bolometric luminosity ($L_{\rm bol}/L_{\rm scale}$) in the $f_{\rm cold}-d_{\rm neigh}/R_{\rm vir,neigh}$ plane for Quasar galaxies. $f_{\rm cold}$ is the fraction of cold gas ($T< \rm 10^4 \ K$) mass in the host galaxies. The red points show the distribution of AGNs in the plane. For galaxies that are cold gas-rich ($log \ f_{\rm cold}>-2$)for ($d_{\rm neigh}/R_{\rm vir,neigh}\leq 0$)  the contours are vertical showing that AGN activity is determined by the cold gas fraction content of host galaxies in the quasar mode. For the host galaxies not rich in cold gas the contours are horizontal, suggesting that environmental effects are important in determining the AGN activity.} \label{fig:gasenv} 
	
\end{figure}

As shown in the analysis above, distance to the neighbor and cold gas content of the host galaxies are shown to affect the AGN activity. We note that the cold gas content must be closely related to the interaction (especially at high redshifts) with neighbors and mergers as satellites will bring new cold gas to the central. A direct comparison of the two therefore should be done in light of this information. We compare these the distance to the neighbour ($d_{\rm neigh}/R_{\rm vir,neigh}$) and the fraction of cold gas mass ($f_{cold}=M_{\rm cold}/M_{\rm gas}$) in Figure \ref{fig:gasenv}. In the region where environmental effects become important  ($log \ d_{\rm neigh}/R_{\rm vir,neigh}\leq 0$) for galaxies which are cold gas-rich ($log \ f_{\rm cold}>-2$) the contours are vertical showing that AGN activity is determined by the cold gas fraction of host galaxies in the quasar mode. For the galaxies not rich in cold gas mass the contours are horizontal, showing that environmental effect becomes important in determining the AGN activity.

\begin{figure} 
	
	\centering
	
	\includegraphics[width=0.45\textwidth]{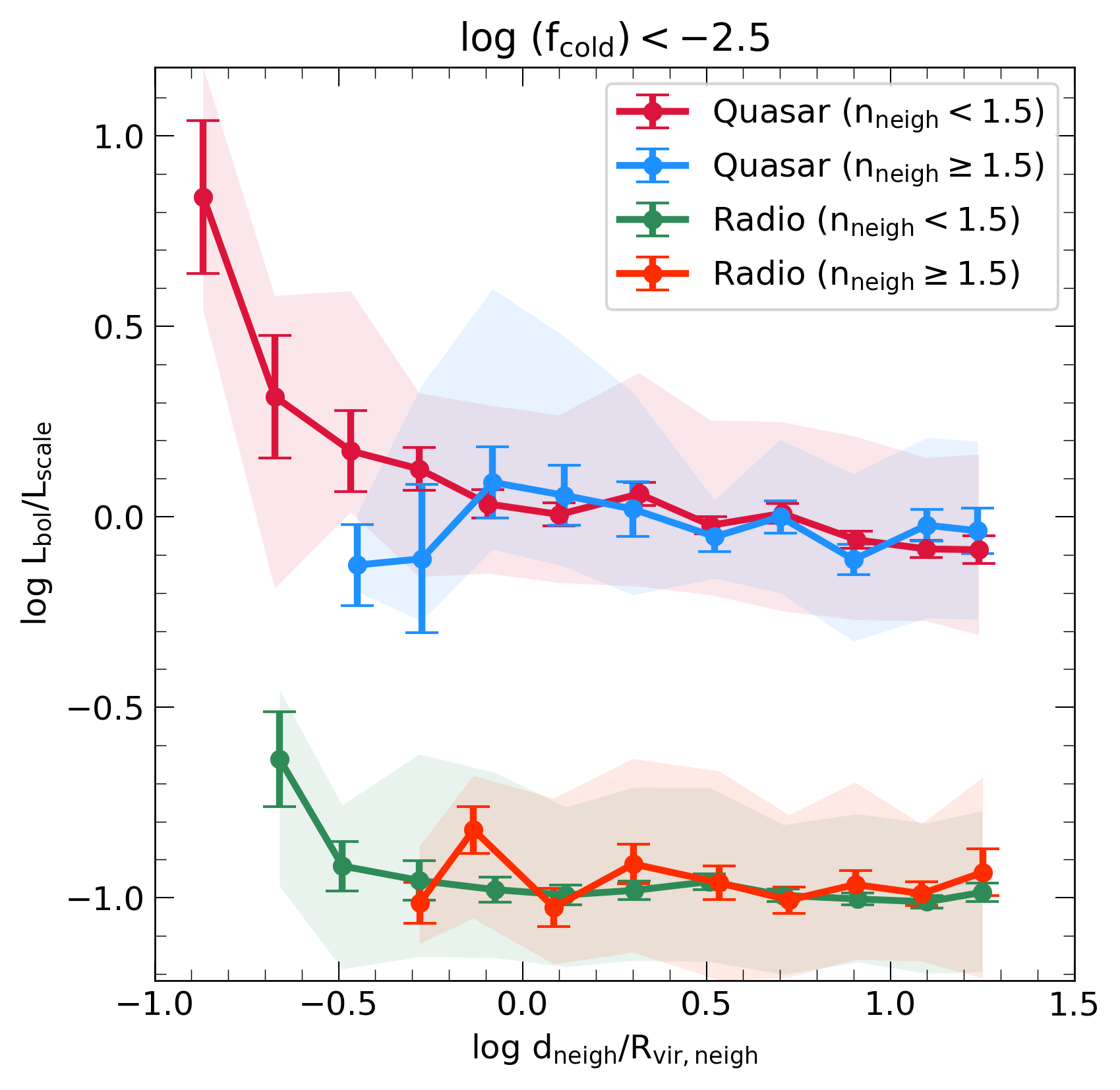}
	
	\caption{Bolometric luminosity of AGNs ($L_{\rm bol}/L_{\rm scale}$)  with total cold gas ($T< \rm 10^4 \ K$) with distance from the neighbor ($d_{\rm neigh}/R_{\rm vir,neigh}$). For visualization, we shift the radio mode values by 1 dex in y-axis. Quasar and radio mode with a late-type neighbor ($n_{neigh}<1.5$) show an increase in  $L_{\rm bol}/L_{\rm scale}$ with a decrease in the distance to the neighbor within one virial radius by 1 and $<0.5$ dex respectively. For early-type morphology ($n_{neigh}\geq1.5$) there is no clear trend.} \label{fig:neighser} 
	
\end{figure}

In Figure  \ref{fig:neighser} we further explore the effect of the environment in the gas-poor galaxies with $log f_{cold} < -2.5$ in the host galaxy of AGNs. We show the changes in the bolometric luminosity ($L_{\rm bol}/L_{\rm scale}$) with a decrease in distance to the neighbor ($d_{\rm neigh}/R_{\rm vir,neigh}$)  for different morphology of the nearest neighbor. For visualization, we shift the radio mode values by 1 dex in y-axis. 
It is evident that for both modes with the late-type neighbor, with  S\'{e}rsic index ($n_{\rm neigh})<1.5$ the bolometric luminosity increases with a decrease in the distance to the neighbor within one half virial radius. The increase is 1 dex for quasar-mode galaxies and less than 0.5 dex for radio-mode galaxies. For an early-type morphology ($n_{\rm neigh}\ge 1.5$) of neighbor there is no clear trend as the distance to neighbor decreases. Therefore, we conclude that although it does not change the overall trend with the environment, the morphology of the neighbor is also important in regulating the AGN activity.

Our interpretation of interaction playing a role in the quasar mode agrees with the previous studies \citep{2005Springel,2005Matteo}. \cite{2005Kawata} who concluded that the AGN phase is triggered and regulated by the gas inflow in the nuclear region of the galaxies using high-resolution simulation. The results discussed here also match the results from \cite{2020Bhowmick}, who used IllustrisTNG simulation and concluded that interaction increases the AGN activity but plays a minor role. Our results complement the findings of \cite{2013Trump} who reported that more luminous AGNs are associated with blue star-forming galaxies and concluded that the process fuels AGN and results in efficient star formation. Our results agree well with \cite{2011Ellison} who showed that AGN activity starts well before the mergers. \cite{2014Satyapal} also found that merger can increase the activity of the AGNs detected in infrared wavelengths. \cite{2021Duplancic} studied AGNs in pairs, triplets, and groups. They found twice the fraction of powerful AGNs residing in pairs and triplets than in regular AGNs. They reported that distance to the neighbor increases the AGN fraction in triplets but does not change it for galaxies in pairs. Our results complement these findings that the effect of a near neighbor would make AGNs more luminous and detectable in the observations.

\cite{2019Man} using SDSS concluded that internal secular evolution is the dominant mechanism for AGN triggering and environment plays a minor role. At a low redshift, \cite{2021Smethurst} studied outflows from the AGNs to test the importance of mergers in triggering the AGNs. They concluded that the non-merger-driven processes (internal processes) are sufficient for triggering the AGNs. 
We find a similar trend in our analysis presence of cold gas in the AGN host galaxy is the dominant property, and the environment plays a role in funneling this gas toward the central black hole. We find complementary results to  \cite{2013Sabater}, who studied the environment of optical and radio galaxies. The analysis showed low nuclear activity in AGNs residing in high-density environments due to a lack of gas supply. Optical AGNs require the presence of cold gas, and one-to-one interaction enhances the presence of AGNs. They also found that radio mode AGNs are triggered by gas cooling from the environment.

\cite{2012Treister} studied AGNs in the luminosity range  $\rm 10^{43}-10^{46} \ erg \ s^{-1}$ in the redshift range ($0<z<3$). They reported at higher redshift ($z>2$), merger-triggered AGNs play a significant role in the growth of central black holes, whereas at low redshifts secular processes become important. For the low redshifts, they concluded the availability of gas plays a crucial role in determining the black hole growth. Their study matches the luminosities of AGNs in HR5 and supports our conclusions. 

Recently, \cite{2022Uchiyama} studied the evolution environment around radio galaxies in the redshift range $z=0.3-1.4$. They estimate the environment of galaxies with the k-neighbor density method. They reported that low mass radio galaxies  ($M_{*}<10^{11} \ M_{\odot}$) reside in similar environment (overdensity $\sim1$) as the control sample. The projected distance to the neighbor was similar for less massive radio galaxies compared to the control sample. They concluded that massive radio mode galaxies grew by experiencing mergers at redshift $z>1.4$ whereas low mass radio galaxies grew by accretion. For the redshift and stellar mass range considered in our study, these results match well with our finding that radio mode AGNs grow by hot-mode accretion and environment plays a minor role.

Thus, with good statistical samples in all the environments and the study of AGNs in the two modes, this study thus provides a unique vantage point to compare with observations (which suffer from incompleteness) and simulations (which suffer from sample size). Our results favor the scenario of internal properties being the dominant factor in controlling nuclear activity, and the environment starts playing a role within a distance of half virial radius of the AGN host galaxy.

\section{Summary }\label{sec:sum}

In this work, we explored the effect of the local environment on the properties of AGN host galaxies in Horizon Run 5. We classify AGNs into radio and quasar mode galaxies based on their Eddington ratio. We study the effects on AGN activities of the intrinsic properties of host galaxies, background density, and distance to neighbor. \par

We demonstrated through our analysis that the AGN activity in quasar mode is correlated with the star formation rate and stellar metallicity of the host galaxy. The star formation rate is the property that is strongly correlated with the bolometric luminosity. The radio mode on the other hand shows much less variation in the AGN activity with these intrinsic parameters. Comparing various internal properties that correlate with AGN activity in quasar mode AGN, we demonstrated that the star formation rate has the strongest correlation. \par

We studied the effect of the environment using two parameters that capture the environment at two scales: the background density and distance to the neighbor. We found that the AGN activity of the galaxy shows an increase with both proximity to a neighbor and higher background density. Total gas mass, on the other hand, shows a decrease with the increase in background density and proximity to a neighbor. Star formation in quasar mode follows the same pattern of increase. Comparing the two environment parameters, we demonstrated proximity to the neighbor is the dominant parameter regulating the AGN activity and star formation rate in quasar mode. Interaction with neighbors also determines the gas content and kinematics of the host galaxies of AGN in both modes. \par

We found that the AGN activity in quasar mode relates to the gas content inside the host galaxy. The quasar mode AGN contains intrinsically more gas than the radio mode AGN. The cold gas content gas inside the 3kpc cut of the central black hole in the quasar mode is significantly more than the radio mode galaxies. We demonstrate that interaction with the environment helps the gas in the host galaxy to lose angular momentum and to fall towards the center in quasar mode triggering AGN activity more efficiently. We found that radio mode AGNs contain more hot gas than quasar mode. The fraction of radio mode in the high-density environment also decreases. \par

Our results support the scenario that the internal parameters, such as gas content, govern the AGN activity. Cold gas content inside the vicinity of the supermassive black hole predominantly governs the AGN activity in quasar mode, whereas radio mode is supported by gas accretion of hot gas inside the host halo.  The environment becomes important in the quasar mode AGN by funneling the cold gas toward the central black hole when the nearest neighbor is within a distance of half the virial radius. 

With the availability of new observations in the future, it will be interesting to compare these results with a statistically good sample at different redshifts and study the evolution of AGN in the Universe. Interaction with the neighbor makes the total gas inside the AGN get consumed. A fraction of gas is converted to stars, and the other is accreted on the central black hole powering the AGN. This fraction of gas varies for different galaxies and is a function of distance to the neighbor. Tracing the evolution of AGN undergoing an interaction in HR5 may reveal more details on this process, and we reserve this for future work.

\acknowledgments

The authors thank the Korea Institute for Advanced Study for providing computing resources (KIAS Center for Advanced Computation Linux Cluster System) for this work, AS and CBP were supported by KIAS Individual Grants (PG080901 and PG016903). JK was supported by a KIAS Individual Grant (KG039603) via the Center for Advanced Computation at Korea Institute for Advanced Study. JL is supported by the National Research Foundation of Korea (NRF-2021R1C1C2011626). YK is supported by the National Research Foundation of Korea (NRF-2020R1C1C1007079). This work benefited from the outstanding support provided by the KISTI National Supercomputing Center and its Nurion Supercomputer through the Grand Challenge Program (KSC-2018-CHA-0003 and KSC-2021-CHA-0012). Large data transfer was supported by KREONET, which is managed and operated by KISTI. BKG and CGF acknowledge the support of STFC through the University of Hull Consolidated Grant ST/R000840/1, access to {\sc viper}, the University of Hull High Performance Computing Facility, and the European Union’s Horizon 2020 research and innovation programme (ChETEC-INFRA -- Project no. 101008324).

		\bibliography{library_agn}

	\end{document}